\documentclass[format=acmsmall, review=false, screen=true]{acmart}
\pdfoutput=1
\usepackage{booktabs} % For formal tables

\usepackage{framed}
\usepackage{hyperref}
\usepackage{balance}
\usepackage{natbib}
\usepackage{epsfig}
\usepackage{color}
\usepackage{subfigure}
\usepackage{multirow,tabularx}

\usepackage{amssymb}
\usepackage{mathtools}
\usepackage{graphicx}
\usepackage{epstopdf}
\usepackage{bm}

\usepackage{csquotes}
\usepackage{array}
\usepackage[yyyymmdd,hhmmss]{datetime}
\usepackage[subject={Todo}]{pdfcomment}
\usepackage[textsize=scriptsize,bordercolor=black!20]{todonotes}
\usepackage{xcolor,colortbl}
\usepackage{amssymb}% http://ctan.org/pkg/amssymb
\usepackage{pifont}% http://ctan.org/pkg/pifont
\usepackage[algo2e,titlenumbered,ruled]{algorithm2e} 
\usepackage{lipsum,environ,amsmath}
\usepackage{slashbox,booktabs,amsmath}

\usepackage{xspace}

\newcounter{Lcount}
\newcommand{\numsquishlist}{
   \begin{list}{\arabic{Lcount}. }
    { \usecounter{Lcount}
 \setlength{\itemsep}{-.1ex}      \setlength{\parsep}{0ex}
      \setlength{\topsep}{0ex}       \setlength{\partopsep}{0ex}
      \setlength{\leftmargin}{1em} \setlength{\labelwidth}{1em}
      \setlength{\labelsep}{0.1em} } }
\newcommand{\numsquishend}{\end{list}}

\newcommand{\squishlist}{
   \begin{list}{$\bullet$}
    { \setlength{\itemsep}{-.1ex}      \setlength{\parsep}{0ex}
      \setlength{\topsep}{0ex}       \setlength{\partopsep}{0ex}
      \setlength{\leftmargin}{.8em} \setlength{\labelwidth}{1em}
      \setlength{\labelsep}{0.5em} } }
\newcommand{\squishend}{\end{list}}

\definecolor{Gray}{gray}{0.85}
\definecolor{LightGray}{rgb}{0.9,0.9,0.9}
\definecolor{LightBlue}{rgb}{0.8,0.8,1}

\newcommand{\clip}{{\sc Coordination Initiator Inference Problem}\xspace}%

\newcounter{problem}
\newenvironment{problem}[1][htb]
  {\renewcommand{\algorithmcfname}{Problem}% Update algorithm name
   \begin{algorithm2e}[#1]%
   \SetAlFnt{\small}
    \SetAlCapFnt{\small}
    \SetAlCapNameFnt{\small}
    \SetAlCapHSkip{0pt}
  }{\end{algorithm2e}}
  
  \newenvironment{alprocedure}[1][htb]
  {\renewcommand{\algorithmcfname}{Procedure}% Update algorithm name
   \begin{algorithm2e}[#1]%
    \SetAlFnt{\small}
\SetAlCapFnt{\small}
\SetAlCapNameFnt{\small}
\SetAlCapHSkip{0pt}
\IncMargin{-\parindent}
   
  }{\end{algorithm2e}}

\setcopyright{acmcopyright}
\acmJournal{TKDD}
\acmYear{2018} \acmVolume{12} \acmNumber{5} \acmArticle{53} \acmMonth{6} \acmPrice{15.00}\acmDOI{10.1145/3201406}

\received{September 2017}
\received[revised]{January 2018}
\received[accepted]{March 2018}

\begin{document}

% Title portion. Note the short title for running heads 
\title{Coordination Event Detection and Initiator Identification in Time Series Data}

\author{Chainarong Amornbunchornvej}
\orcid{0000-0003-3131-0370}
\affiliation{
\institution{University of Illinois at Chicago}
  \streetaddress{851 S Morgan St}
  \city{Chicago}
  \state{IL}
  \postcode{60607-7101}
  \country{USA}}
\email{camorn2@uic.edu}

\author{Ivan Brugere}
\affiliation{
\institution{University of Illinois at Chicago}
  \streetaddress{851 S Morgan St}
  \city{Chicago}
  \state{IL}
  \postcode{60607-7101}
  \country{USA}}
\email{ibruge2@uic.edu}

\author{Ariana Strandburg-Peshkin}
\affiliation{
\institution{Max Planck Institute for Ornithology}
  \streetaddress{Am Obstberg 1}
  \city{Radolfzell}
  \postcode{78315}
  \country{Germany}}
\email{astrandburg@orn.mpg.de}

\author{Damien R. Farine}
\affiliation{
\institution{Max Planck Institute for Ornithology}
  \streetaddress{Universitätsstrasse 10}
  \city{Konstanz}
  \postcode{78464}
  \country{Germany}}
\email{dfarine@orn.mpg.de}

\author{Margaret C. Crofoot}
\affiliation{
\institution{University of California, Davis}  
  \streetaddress{1 Shields Ave}
  \city{Davis}
  \state{CA}
  \postcode{95616}
  \country{USA}}
\email{mccrofoot@ucdavis.edu}

\author{Tanya Y. Berger-Wolf}\thanks{This work was supported in part by the NSF grants III-1514126 (Berger-Wolf, Crofoot), III-1514174 (Crofoot), IOS-1250895 (Crofoot), SMA-1620391 (Crofoot) , CNS-1248080 (Berger-Wolf), and  the David \& Lucile Packard Foundation 2016-65130 (Crofoot). } 
\affiliation{
\institution{University of Illinois at Chicago}
  \streetaddress{851 S Morgan St}
  \city{Chicago}
  \state{IL}
  \postcode{60607-7101}
  \country{USA}}
\email{tanyabw@uic.edu}

\begin{abstract} \small\baselineskip=9pt 
Behavior initiation is a form of leadership and is an important aspect of social organization that affects the processes of group formation, dynamics, and decision-making in human societies and other social animal species. In this work, we formalize the  \clip and propose a simple yet powerful framework for extracting periods of coordinated activity and determining individuals who initiated this coordination, %as well as the mechanism of coordination,
based solely on the activity of individuals within a group during those periods. 
The proposed approach, given arbitrary individual time series, automatically (1) identifies times of coordinated group activity, (2) determines the identities of initiators of those activities, and (3) classifies the likely mechanism by which the group coordination occurred, all of which are novel computational tasks. We demonstrate our framework on both simulated and real-world data: trajectories tracking of animals as well as stock market data. Our method is competitive with existing global leadership inference methods but provides the first approaches for local leadership and coordination mechanism classification. Our results are consistent with ground-truthed biological data and the framework finds many known events in financial data which are not otherwise reflected in the aggregate NASDAQ index. Our method is easily generalizable to any coordinated time-series data from interacting entities. 
\end{abstract}

%
% The code below should be generated by the tool at
% http://dl.acm.org/ccs.cfm
% Please copy and paste the code instead of the example below. 
%



% The default list of authors is too long for headers}
\renewcommand{\shortauthors}{C. Amornbunchornvej et al.}

\maketitle

\section{Introduction}

 Which zebra initiated the flight from a lion? Whom does the elephant herd follow to water? Who is the trend-setter whose opinion many follow at the moment? (And is it the same person whether it's the opinion about the future of AI or the hottest lunch spot?) In all these scenarios, the initiator might not be the one who is speaking the loudest or positioned at the front of the group \textit{after} the group has already agreed to follow~\cite{Dyer:2009aa,Stewart:1947aa}. Thus, in order to identify those initiators or trend-setters, we must also determine the moment of the group's decision to follow. 
\begin{framed}
\noindent {\clip:} {An agreement of a group to follow a common purpose is manifested by its coalescence into a \textit{coordinated} behavior. The process of initiating this behavior and the period of decision-making by the group members necessarily precedes the coordinated behavior. {\bf Given time series of group members' behavior, the goal is to find these periods of decision-making and identify the initiating individual, if one exists.}}
\end{framed}

{Initiating} a group's behavior is a form of leadership~\cite{WilsonSocbio,Stueckle2008}. Leadership is an important aspect of the social organization, formation, and decision-making of groups of people in online and offline communities, as well as other social animals. Understanding the dynamics of emerging leadership allows researchers to gain insights into how social species make decisions. Until recently, many works defined leaders by their physical or behavioral characteristics rather than by observing processes of interaction~\cite{LeadershipBook}. %the success of their initiatives. \todo{citation needed}
%it has been difficult if not impossible to pinpoint the identity of a leader from available observational data without explicit additional information. However,

\begin{figure}[ht!]
\centering
\includegraphics[width=1\columnwidth]{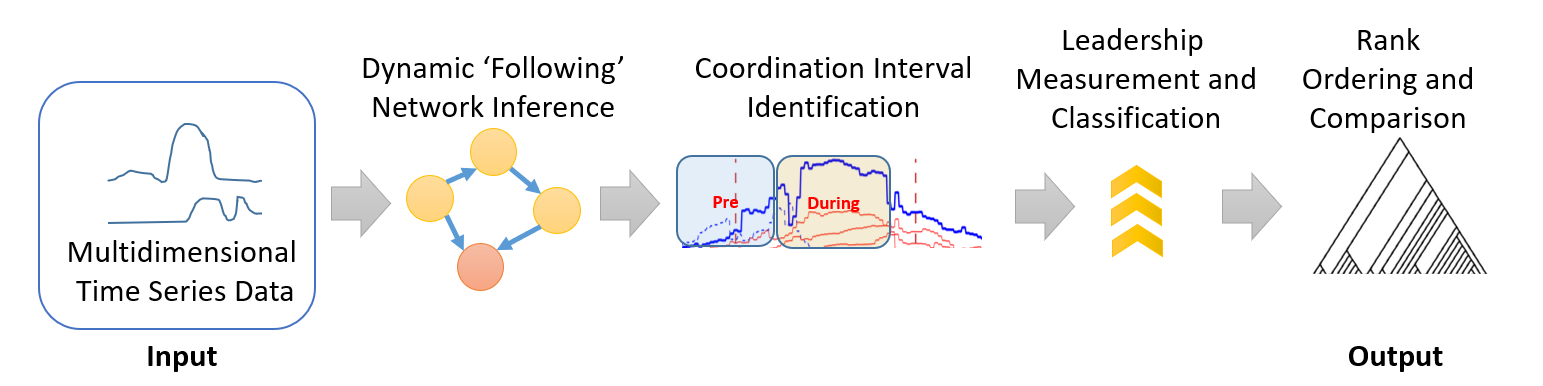}
\caption{A high-level overview of the proposed framework}
\label{fig:processingSteps}
\end{figure}
The availability of data from physical proximity sensors, GPS, and the web opens up the possibility of measuring leadership as the process of initiation in online activities, face-to-face human interactions, animal populations, and aggregate social processes such as economic activity. This paper presents the new computational problem of inferring leader identity in the context of successful initiation of coordinated activities among groups of individuals or other entities, as well as proposes the first automated method for unsupervised leader identification. The method uses only time series activity data of entities, with no additional information. The proposed approach automatically determines (1) the time interval of group coordination, (2) the time when the (possibly implicit)  decision for that coordinated activity was made, (3) the identity of the coordination initiator, and (4) the mechanism by which the group came to follow the initiator. %In the context of initiating time-series coordination, the collective consensus may be passive and implicit rather than active.

\subsection{Related work}
Coordinating patterns of individual activity is a challenge that all social organisms face. Diverse strategies--from democratic to dictatorial--have emerged to allow members of groups to reach consensus~\cite{Conradt:2003aa}. Leadership %(defined as non-random, differential influence~\cite{Smith:2016aa})  
plays a key role in organizing the collective (i.e. group) behaviors of social organisms ranging from humans ~\cite{Dyer:2009aa} 
%to hyaena~\cite{Smith2015187} 
to hymenoptera~\cite{Weinstein:1997}. It potentiates complex patterns of cooperation and conflict 
(e.g.,
 lions~\cite{Heinsohn1260}, 
hyaenas~\cite{Boydston:2001}, 
meerkat~\cite{Mares3989}, chimpanzees~\cite{Gilby:2015}, humans~\cite{Glowacki:2015}), 
%~\cite{Heinsohn1260, Gilby:2015, Glowacki:2015},
organizes group movements 
%(fish~\cite{Couzin:2005aa}, humans~\cite{Dyer:2009aa})
~\cite{Couzin:2005aa,Dyer:2009aa}, and may prevent free-riding~\cite{Hooper2010633}. 

In the context of group behavior and decision-making in biology and sociology, leaders are individuals who successfully induce a group of others to follow them to a common goal, state, or behavior~\cite{WilsonSocbio,Stueckle2008,Couzin:2005aa,Petit2010635}. Biological studies showed that leaders may be context-specific~\cite{Song2014crowdmodel,Couzin:2005aa} and the important initiators of particular group activities are not necessarily the individuals found at the top of their group's social dominance hierarchy~\cite{%Stewart:1947aa,VanVugt01112006,
Brent2015746,Strandburg-Peshkin1358}. 
%They may not even be explicitly known to the rest of the group but may possess prior knowledge that allows them to move towards resources while others follow~\cite{Song2014crowdmodel,Couzin:2005aa}. 

Substantial interest currently exists in identifying leaders and determining how they influence the behavior of others in their social environment.  % \cite{Sun2011}.
Previous work in several domains defined leadership according to physical characteristics ({\em e.g.,} size, sex~\cite{WilsonSocbio}), positions in % physical movement in public spaces \cite{solera2015learning},
 location-based social networks \cite{PhamICDE2016}, rule-based models\cite{Song2014crowdmodel}, physical trajectory and association patterns \cite{lusseau2009emergence,andersson2008reporting}. 

Computationally, most previous  work uses a global notion of leadership and creates a global, static leadership ranking over the entirety of the input data \cite{goyal2008discovering, %Tan:2010:SAT:1835804.1835936%,solera2015learning,
 Bakshy:2011:EIQ:1935826.1935845}. Other domain-specific methods infer leadership from implicit pairwise dyadic dominance or leader-follower interactions~\cite{andersson2008reporting, PhamICDE2016,kjargaard2013time}. Some methods define an explicit network over the dyadic interactions or use a known network topology~\cite{Sun2011} and use network measures, such as PageRank and HITS, or cascade size to identify leaders~\cite{Bakshy:2011:EIQ:1935826.1935845}.

%General approaches which only consider dyadic interactions or relationships between pairs of individuals in isolation will likely be insufficient for unravelling the complex dynamics of consensus building. In small scale human societies, for example, it is social support from both kin and non-kin that gives rise to leadership ability, and centrality within their social network is what characterizes the emergent leaders ~\cite{vonRueden2015978}. 

Leadership has also been studied in explicit social network settings.
From a social network perspective, leaders can be characterized as influential individuals who have many followers that imitate the leader's actions~\cite{goyal2008discovering}, and, thus, successfully take a group from one behavioral state to another. Much of the computational work has focused on the problem of influence maximization (IM)--i.e. how individuals are able to maximize their impact on the behavior of the group as a whole by iteratively affecting local network neighborhoods~\cite{kempe2003maximizing, goyal2010learning}. This approach assumes that the network structure is known.

Alternatively, leadership can be viewed as the causal effect for the followers' actions. Granger Causality~\cite{10.2307/1912791} is one of the methods used for inferring cause and effect within a set of time series. The idea is that if we can use time series $Y$ to significantly improve the prediction of the future activity of time series $X$, compared to using only the past information from $X$,  then $Y$ Granger causes $X$. Certainly, this approach does not distinguish between a direct causality and confounding effects. There is a large body of work using Granger Causality to infer temporal dependency among time series~\cite{Arnold:2007:TCM:1281192.1281203,liu2009nonparanormal,liu2012sparse}.  However, there is no explicit work using Granger Causality to infer leaders in time series, though it is certainly a fruitful direction to explore.  

There is a clear gap between the biosociological definitions of leadership in group decision-making and the existing computational approaches. Currently, there are no computational approaches that (1) view leaders as initiators of group behavior change, which can (2) identify the timing of the process of the change initiation and the group's decision-making in (3) arbitrary contexts, under (4) a variety of leadership models.

\subsection{Our contributions}
In our previous version of this paper in ~\cite{flicaMilets}, we focus on the definition of leadership as the initiation of coordinated activities. We aim to close the gap between the biosociological view of the role of leaders in group decision-making, the computational formalism, and the methodology.

Therefore, the first part of our contribution is establishing and formalizing this {\bf new computational problem of coordination initiation inference}. We call it the \clip. Our formulation is a \textbf{generalization} of many related leadership inference computational problems. We explicitly relate existing leadership and influence propagation problems as special cases of our formulation. The new formulation uses only the time series of individual behavior as input, with no assumption of additional information such as demography, prior historic data, dominance hierarchy, or a network structure. The problem formulation aims to identify different local instances of behavior initiation, allows the identity of the initiator to be instance-specific, and makes no assumption on the leadership or behavioral model.

Our additional contribution is in proposing a computational {\bf solution framework} to this new \clip. We propose a general, scientifically grounded, unsupervised, and extendable framework with few assumptions for identifying individuals who lead a group to a state of coordinated activity (or, more generally, an entity that induces group coalescence). Our framework is capable of:
\squishlist
\item {\bf Detecting coordinated activity events:} discovering coordination intervals and decision-making periods leading to that coordination;
\item {\bf Identifying initiators:} identifying the initiators of this coordinated behavior, that is, the individuals who succeeded in leading the group to coordination, specifically locally to each coordination instance; and
\item {\bf Classifying the group coordination model:} characterizing the type of the group's transition behavior to coordination according to interpretable, dynamic models. %(e.g. hierarchical, dictatorial, local influence).
\squishend

We demonstrate the framework's ability to analyze leadership in coordinated activity on synthetic and real datasets over several domains. We compare our framework with state-of-the-art methods for leadership identification for the special cases of our problem where such methods are applicable. For many instances of our new problem, there are no existing methods. We demonstrate that existing solutions fail and do not extend to these instances. We use synthetic simulated data to validate each aspect of the framework. We analyze two biological datasets -- GPS tracks of a baboon troop and video-tracking of fish schools -- as well as stock market closing price data of the NASDAQ index. The results are consistent with ground-truthed biological data. Moreover, the framework finds many known events in financial data, which are not otherwise reflected in the aggregate NASDAQ index. Our approach is easily generalizable to any coordinated activity in time series data of interacting entities. \\

In addition, in this paper, we propose a new group activity classification method based on the {\em  following network} framework. We use it to classify group activity in a dataset of GPS tracking of a troop of baboons.   There are four group activities in the baboon dataset: sleeping, hanging-out, coordinated non-progression, and coordinated progression. By using only the network density as a feature, our simple classifier performed better than the state-of-the-art classifier method which used 24 features in baboon activities classification task~\cite{li2016adversarial}.  %In addition to previous papers, we provided more details and description of material in this paper.

\subsection{Influence Maximization vs. Coordination Initiator Inference Problem}
The Influence Maximization problem is closely related to \clip. In fact, {\em successful} Influence Maximization, where a large fraction of the population is influenced, is a special case of \clip. When the influence is spread to a majority of the population, that population is now in a coordinated state, with the decision period starting at the initiation of the influence and the initiators being that source of influence. However, \clip goes beyond Influence Maximization in every aspect of the framework and can capture different models of decision-making, coordinated activity, as well as repeated and context-specific dynamics of coordination. 
\squishlist
\item {\bf Coordination Model:} In Influence Maximization, the majority of papers focus on Linear Threshold and Independent Cascade models as main coordination mechanisms. However, there are other models that can be represented as a coordination mechanism, such as Dictatorship and Hierarchy models, as well as non-network based models at all. Our new problem formulation, \clip, generalizes to all types of models that can be represented as a coordination mechanism. To recognize difference types of coordination mechanisms, we also provide the model classification approach to classify these coordination models based on some proposed features.
\item {\bf Coordinated Activity:} Influence Maximization focuses on only one kind of state changing, which is an information spreading event. In a coordinated activity, any state that the group coalesces to, a posteriori, is a coordinated state, and it may change from one time to another. Coordinated activity can represent multiple types of coalesced states over multiple time points. For example, a coordinated movement of animals to a feeding site is considered to be a coordinated activity for that group, but so is sudden jumping up and down in agitation, or all looking in the same direction, or all falling sick. And the coordinated movement can be different and look different in the morning versus in the evening. In \clip, the problem focuses on not only which individuals initiate coordinated activities, but also {\bf when} coordinated activities occur, without the explicit prescription of the type of coordinated activity. To analyze coordinated activities in time series, we provide a framework to detect coordination intervals as well as initiators of these coordination events. 
\item {\bf Coordination Based on Context:} The original Influence Maximization assumes that there is only one coordination event of information spreading.  However, in reality, information spreading or coordination events can occur many times. For instance, within a day, a group of animals can have many coordinated movement events to many places. Our work here aims to infer the multiple coordination events, which might have different initiators.
\squishend

\newcommand{\argmax}{\mathop{\mathrm{argmax}}\limits} 
\section{Problem formalization}

Given a collection of time series, we want to find initiators of highly coordinated patterns. To formally state the \clip, we need to formalize notions of        ~``coordination'' and ``initiation.'' 

%Given a pair of time series, we introduce the concept of $t$-shift and coordination in order to define an initiator as follows. 

\begin{figure}[ht!]
\centering
\includegraphics[width=.75\columnwidth]{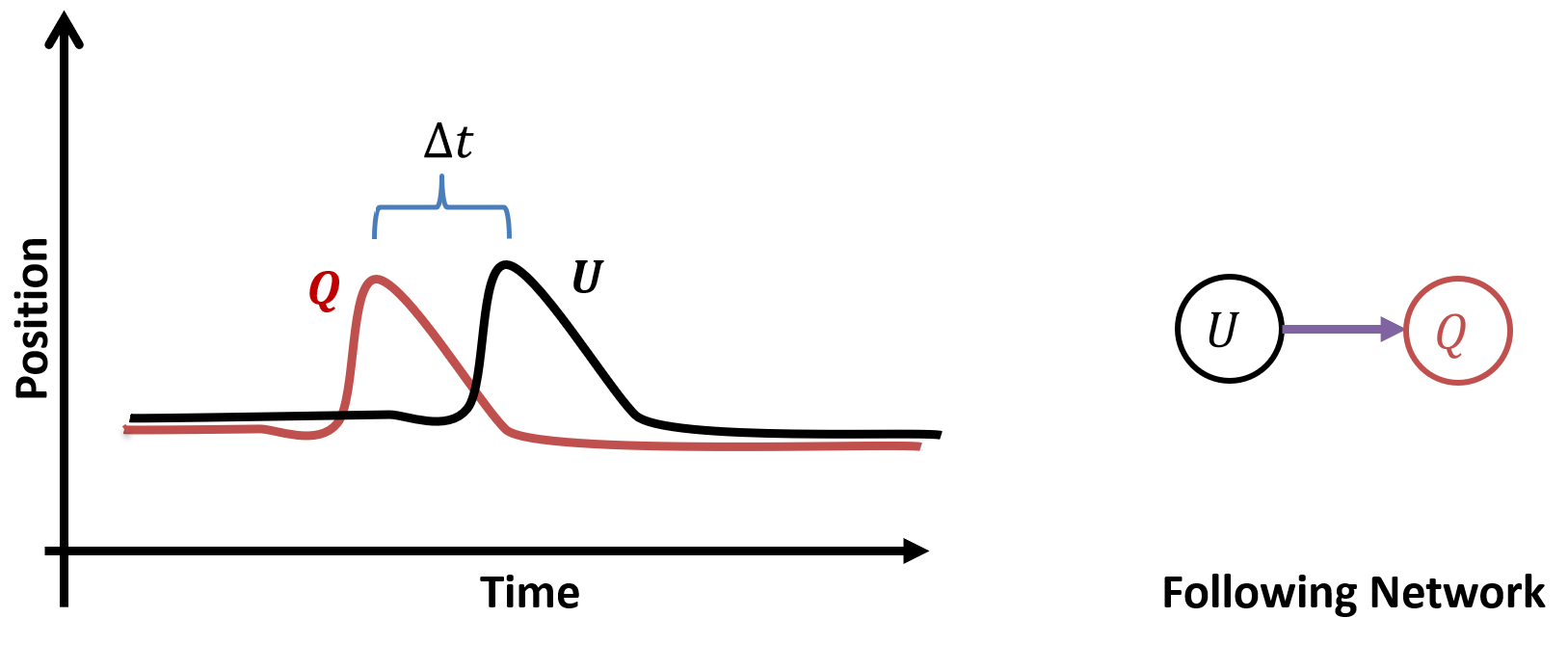}
\caption{(Left) The example of time series $U$ follows $Q$. (Right) the following network w.r.t. $U$ follows $Q$ relation.}
\label{fig:FollRelation}
\end{figure}

First, we define an intuitive notion of a {\sc following relation}, as ``two individuals performing the same sequence of actions (or generating time series values) with some fixed delay (Fig.~\ref{fig:FollRelation}).'' Formally:

\begin{definition}[{\sc following relation}]
\label{followRDef}
Let $U = (\vec{u}_1,\dots,\vec{u}_t,\dots)$ and $W = (\vec{w}_1,\dots,\vec{w}_t,\dots)$ be $m$-dimensional, arbitrary-length time series. If for all $ t \in \mathbb{N}$, there is a fixed time delay $\Delta t \in \mathbb{Z^+} \cup\{0\}$ such that $\vec{w}_t=\vec{u}_{t+\Delta t}$
%$U$ and $W$ have maximum similarity $\mathrm{sim}( (\vec{w}_{1},...),(\vec{u}_{1+\Delta t},...) ) \ge \sigma$ with a time delay $\Delta t \in \mathbb{Z^+} \cup\{0\}$
, then $U$ follows $W$ denoted as $W \preceq U$. We denote $W \prec U$ if $\Delta t>0$. 
\end{definition}

\begin{lemma}
\label{EquivR}
Let $U$ and $W$ be time series such that $W \preceq U$ and $U \preceq W$, then $U$ and $W$ are equivalent time series denoted $U \equiv W$.
%Let $U$ and $W$ be time series such that $W \preceq U$ and $U \preceq W$, then $U$ and $W$ are equivalent time series denoted $U = W$.
\end{lemma}
\begin{proof}
There are two cases when both  $W \preceq U$ and $U \preceq W$. First,  $W = U$ and $U = W$ (that is, $\Delta t = 0$ in both following relations). Clearly, $W \equiv U$.
Second, $W \prec U$ with $\Delta t_w > 0$ and $U \prec W$ with $\Delta t_u > 0$. Then, by definition, $\vec{w}_t=\vec{u}_{t+\Delta t_w}$ and $\vec{u}_t=\vec{w}_{t+\Delta t_u}$. Therefore, $\vec{w}_t=\vec{w}_{t+\Delta t_w+\Delta t_u}$. Thus, if  $W \preceq U$ and $U \preceq W$ then $W \preceq W$ (and similarly $U \preceq U$).  Thus, the two time series are identical periodic with a different starting point and therefore equivalent.
%(Thus, since we assumed the relation is symmetric and is trivially reflexive and transitive, it is therefore, by definition, is an equivalence relation.)

%First, $U \equiv U$ since $U \preceq U$ with $\Delta t=0$ (reflexivity). Second, $U \equiv W$ if and only if $W \equiv U$ since both terms are similar (symmetry). Third, if $U \equiv W$ and $W \equiv Q$, then $U \equiv Q$. Since  $U \preceq W$ and $W \preceq Q$, then $U \preceq Q$.  Similarly, $W \preceq U$ and $Q \preceq W$, hence $Q \preceq U$. Therefore, $U \equiv Q$ (transitivity).
\end{proof}

\begin{figure}[ht!]
\centering
\includegraphics[width=.75\columnwidth]{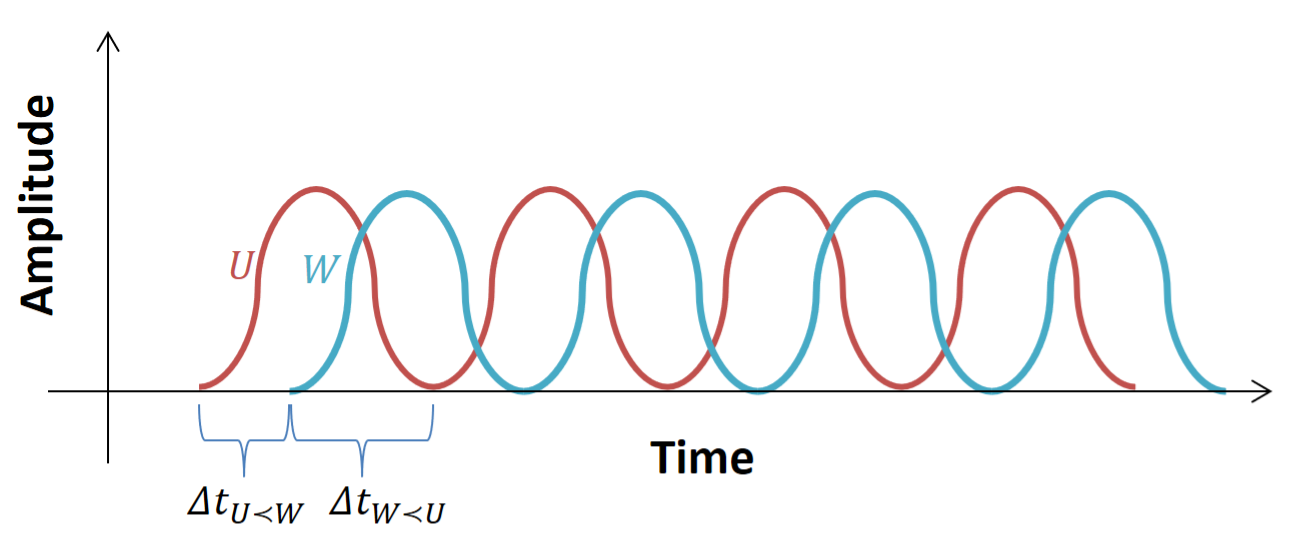}
\caption{$U$ and $W$ are sine waves that have the same frequency but different phase. $U$ follows $W$ with time delay $\Delta t_{W \prec U}$ and $W$ follows $U$ with time delay $\Delta t_{U \prec W}$.}
\label{fig:sineTS}
\end{figure}

In periodic time series, such as a sine wave, we use Lemma~\ref{EquivR}. If we have two sine waves that have the same frequency but different phase, then we consider them as a pair of time series that follow each other, from which it can be concluded that they are equivalent.   For instance, in Fig.~\ref{fig:sineTS}, $U$ and $W$ are two sine waves, which have the same frequency but different phase. Because $U \prec W$ with time delay $\Delta t_{U \prec W} > 0$ and $W \prec U$ with $\Delta t_{W \prec U} > 0$ , by the second case in Lemma~\ref{EquivR} above,  $U \equiv W$.

\begin{lemma}
%Let $\mathcal{U} = \{U_1, \dots, U_n\}$ be a time series set with a following relation over $\mathcal{U}$. The set $\mathcal{U}$ is a partial order set~\cite{partialOrder}.
The {\sc following relation} is a partial order over time series~\cite{partialOrder}.
\end{lemma}
\begin{proof}
%Reflexivity: $\forall U, \; U\preceq U$.
Antisymmetry: if $W\preceq U$ and $U\preceq W$, then $W \equiv U$ by Lemma~\ref{EquivR}.
%Transitivity: if $W\preceq U$ with a time delay $\Delta t_w$ and $U\preceq Q$ with a time delay $\Delta t_u$, then, by definition $W\preceq Q$  with a time delay $\Delta t_w + \Delta t_u$. 
The {\sc following relation} is also trivially reflexive and transitive, which, by definition is a partial order.

\end{proof}

\begin{figure}[ht!]
\centering
\includegraphics[width=.75\columnwidth]{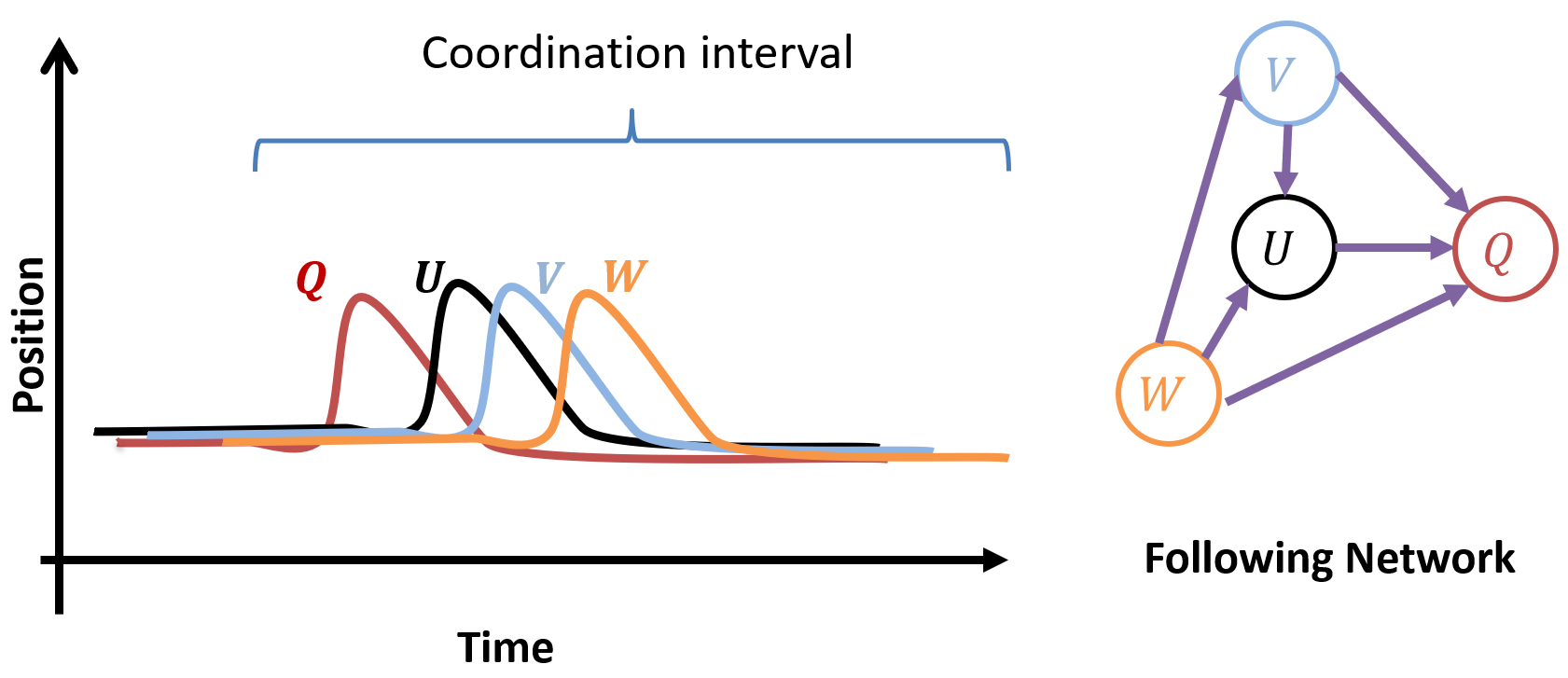}
\caption{(Left) the example of coordination interval in time series where $Q\prec U\prec V \prec W$. (Right) the following network w.r.t. these following relations. In this example, $Q$ is an initiator.}
\label{fig:CoorInterval}
\end{figure}

Next, {\sc coordination}, or intuitively ``all individuals performing the same sequence of actions, at possibly varying delays (Fig.~\ref{fig:CoorInterval}),'' is formally defined as:

\begin{definition}[{\sc Coordination}]	
\label{CoorDef}
Given a set of $m$-dimensional time series $\mathcal{U} = \{U_1, \dots, U_n\}$. The set $\mathcal{U}$ is {\em coordinated} at time $t$ if for every ${n \choose 2}$ pairs $U_i, U_j \in \mathcal{U}$, either $U_i \prec U_j$ or $U_j \prec U_i$. 
The {\em coordination interval} is the maximal contiguous time interval $[t_1, t_2]$ such that $\mathcal{U}$ is coordinated for every $t \in [t_1, t_2]$.
\end{definition}

Finally, the {\sc initiator} is intuitively ``an individual who first performs a sequence of actions, and all other individuals follow,'' formally defined as:

\begin{definition}[{\sc Initiator}] 
\label{InitiatorDef}
Let $\mathcal{U} = \{U_1, \dots, U_n\}$ be a coordinated set of  $m$-dimensional time series within some coordination interval $[t_1, t_2]$. Then the time series $L \in \mathcal{U}$ is the {\em initiator}  time series for the coordination interval if for each time series $U \in \mathcal{U} \setminus \{L\}$, $L \prec U$. 
\end{definition}

In Fig.~\ref{fig:CoorInterval}, $Q$ is an initiator of coordination. The coordination interval starts at the beginning of $W$. We are now ready to precisely state the problem of identifying the individual who initiates a coordinated behavior:

%Now, given those formal definition, we can formally state the problem of identifying the initiator for coordination interval.

\begin{problem}[h!]
    \SetKwInOut{Input}{Input}
    \SetKwInOut{Output}{Output}
    \Input{Set  $\mathcal{U} = \{U_1, U_2, \dots, U_n\}$ of time series.}
    \Output{A coordination interval $[t_1, t_2]$ and the initiator time series $L\in \mathcal{U}$ that initiated the coordination.}
    \caption{{\clip}}
	\label{initCoProb}
\end{problem}

%In the next part, we show that a following relation over a time series set is a partial order and we can use PagRank\cite{PageRank_Brin1998107} to find the following order of time series.

\subsection{Useful observations}
\label{sec:usefulobs}
Let $\mathcal{U}$ be a coordinated set of time series and $L \in \mathcal{U}$ be the initiator. Since $\mathcal{U}$ is a partial order set and $\forall U_i \in \mathcal{U},\;  L \prec U_i$, then, by definition, $L$ is the minimal element. Moreover, $\mathcal{U}$ is a linear order set since for every pair $U_i,U_j \in \mathcal{U}$,  either $U_i \prec U_j$ or $U_j \prec U_i$.

\begin{definition}[{\sc Following network}]
\label{followNetDef}
Let $\mathcal{U} = \{U_1, \dots, U_n\}$ be a set of time series. The {\em following network} $G = (V,E)$ is defined as a directed graph where the set of nodes $V$ has a one-to-one correspondence to the set of time series $\mathcal{U}$, and each edge in $E$ represents a {\sc following relation} between two  time series: $\forall U_i, U_j \in \mathcal{U}$ the edge $e_{i,j} \in E$ if $U_j \prec U_i$.
\end{definition}

Recall that PageRank~\cite{PageRank_Brin1998107} score, $\pi_i$, of a node $i$ in a  network $G$ is defined as follows: 

\begin{equation}
\pi_i=d\sum_{k \in \mathcal{N}^{in}_i} e_{k,i}\pi_k/| \mathcal{N}^{out}_k|+(1-d)
\label{eq:PReq}
\end{equation}

Where $\pi_i \in [0,1]$, \; $d\in (0,1]$ is a constant number, $e_{k,i} \in \{0,1\}$ is one if $e_{k,i} \in E$, $\mathcal{N}^{in}_i$ is a set of neighbor nodes of $i$ such that $k \in\mathcal{N}^{in}_i$ if $e_{k,i} \in E$, and $\mathcal{N}^{out}_i$ is a set of outgoing neighbor nodes of $i$ such that $k \in\mathcal{N}^{out}_i$ if $e_{i,k} \in E$.

\begin{lemma}
\label{PageRankLemma}
Let $G=(V,E)$ be a following network of time series set $\mathcal{U} = \{U_1, \dots, U_n\}$. If $U_i \preceq U_j$ then $\pi_i \geq \pi_j$.
\end{lemma}
\begin{proof}
By transitivity, if $U_j$ follows $U_i$ then the followers of $U_j$ are also the followers of $U_i$. Thus, since $\forall k \in \mathcal{N}_j$,  $U_j \preceq U_k$ and $U_i \preceq U_j$, then $\mathcal{N}^{in}_j \subseteq \mathcal{N}^{in}_i$. Hence, $\pi_i - \pi_j = d\sum_{k \in \mathcal{N}^{in}_i \setminus \mathcal{N}^{in}_j}  e_{k,i}\pi_k/| \mathcal{N}^{out}_k|  \geq 0$.  
\end{proof}
As a corollary of Lemma~\ref{PageRankLemma}, since all the time series follow the initiator $L$ within the coordination interval $[t_1,t_2]$, then $L$ has the highest PageRank score in ${\mathcal U}$ during that coordination period. Moreover, Lemma~\ref{PageRankLemma} allows us to infer the order of following among the time series within the coordination period, as defined by the PageRank values.

\subsection{Following relation with noise}

\begin{figure}[ht!]
\centering
\includegraphics[width=.75\columnwidth]{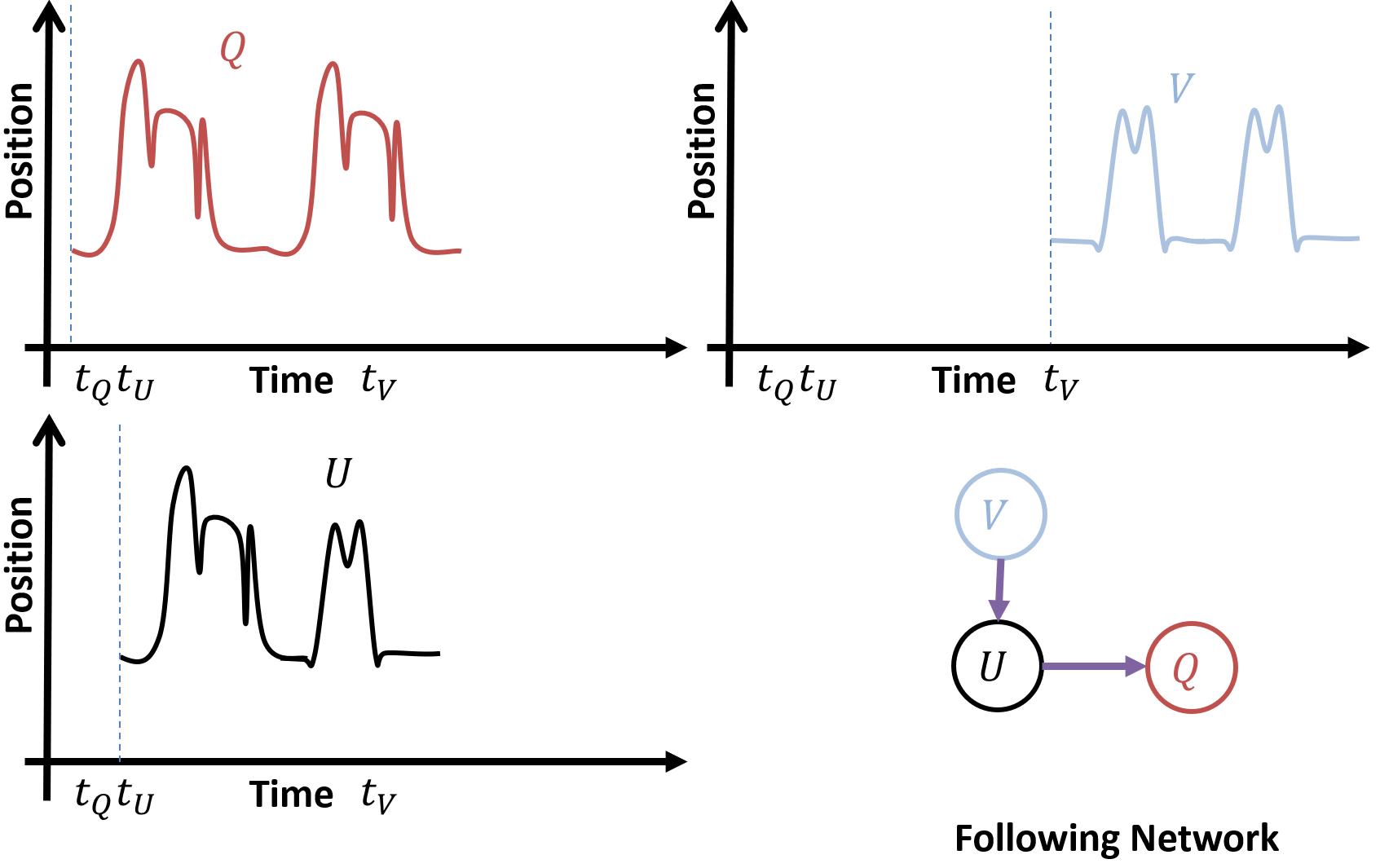}
\caption{(Top-left) A time series $Q$, (bottom-left) a time series $U$, and (top-right) a time series $V$. In this example, $U$ partially follows $Q$, $V$ partially follows $U$ but $V$ does not follow $Q$. The following network of these time series is at the bottom-right of the figure.)}
\label{fig:RelaxTransv}
\end{figure}

In real situations, Def.~\ref{followRDef} requires the exact match, which rarely happens. Therefore, we provide relaxation of following relation to deal with noise in realistic situations below.

\begin{definition}[$\sigma$-following relation] 
\label{def:sigFollRe}
Let $\mathcal{U}$ be a set of time series  and $\mathrm{sim}: \mathcal{U}\times \mathcal{U} \to [0,1]$ be some similarity measure between two time series. For any pair of time series $U_i,U_j \in \mathcal{U}$, let $\Delta t_{max} =\mathrm{min} \argmax_{\Delta t \in \mathbb{Z}} \mathrm{sim}(U_{i,1},U_{j,1+\Delta t})$ where $U_{i,t} \in \mathcal{U}$ represents a time series $U_i$ starting at time $t$, and let $\mathrm{sim}_{max}(U_i,U_j)= \mathrm{sim}(U_{i,1},U_{j,1+\Delta t_{max}})$. Then, for a threshold $\sigma \in (0,1]$, if $ \mathrm{sim}_{max}(U_i,U_j) \geq \sigma$, then we have: 
\squishlist 
  \item if $\Delta t_{max} >0$ , then ${U_i \prec_{\sigma} U_j}$,
  \item if $\Delta t_{max} <0$ , then ${U_j \prec_{\sigma} U_i}$,
	\item if either $\Delta t_{max} = 0$ or ${U_i \prec_{\sigma} U_j}$ and ${U_j \prec_{\sigma} U_i}$, then ${U_i \equiv_{\sigma} U_j}$.
\squishend
\label{SigmaFollwDef}
\end{definition}

The difference between a following relation in Def.~\ref{followRDef} and Def.~\ref{def:sigFollRe} is that the notion of $\sigma$-following relation lacks the transitivity property\footnote{This is similar to non-transitive dice: \url{https://en.wikipedia.org/wiki/Nontransitive_dice}}. Fig~\ref{fig:RelaxTransv} shows the example of three time series and their $\sigma$-following relation with some unknown $\sigma>0.5$. In this example, $Q$ is similar to $U$ and $U$ is similar to $V$ greater than 0.5. However, $Q$ and $V$ are not similar at all. Therefore, the $\sigma$ following relation does not possess the transitivity property.\\

Next, we can define a notion of coordination by using $\sigma$-following relation as follows.

\begin{definition}[{\sc $\sigma$-Coordinated set}]	
\label{SigCoorDef}
Given a set of $m$-dimensional time series $\mathcal{U} = \{U_1, \dots, U_n\}$ and a similarity threshold $\sigma \in (0,1]$. The set $\mathcal{U}$ is {\em $\sigma$-coordinated} at time $t$ if for every ${n \choose 2}$ pairs $U_i, U_j \in \mathcal{U}$, either ${U_i \prec_{\sigma} U_j}$ or ${U_j \prec_{\sigma} U_i}$. 
\end{definition}

\begin{definition}[{\sc $\sigma$-Coordination interval}]	
\label{SigCoorIntvDef}
The {\em $\sigma$-coordination interval} is the maximal contiguous time interval $[t_1, t_2]$ such that $\mathcal{U}$ is coordinated for every $t \in [t_1, t_2]$.
\end{definition}

Even though an initiator of $\sigma$-Coordination interval is not a minimum element anymore due to $\sigma$-following relation does not possess transitivity property, an initiator suppose to have a highest number of followers. Hence, PageRank is still an appropriate measure for finding an initiator.  
\section{Methods}
\label{sec:method}
In this section, we present a Framework for Leader Identification in Coordinated Activity (FLICA) as the solution for the \clip. On real data, the above formalization is very restrictive, so we relax the \textit{exact} {\sc following relation}, and \textit{full} {\sc coordination} to identify `following' and partial `coordination' in real applications.
%\footnote{However, FLICA using PageRank necessarily provides an exact solution to the \clip.} 
Furthermore, multiple coordination events often exist within a set of real time series data. Constructing a single aggregated network would not capture the dynamics of these events. Therefore, FLICA uses a dynamic network approach. 

Fig.~\ref{fig:processingSteps} shows the framework overview. At each time step, we infer following relations to construct a sequence of following networks. We then use network density to detect intervals of coordination, and the time series of PageRank values to identify the initiators of these coordination intervals. 

\begin{figure}[ht]
\centering     %%% not \center
\subfigure{\label{fig:baboon-exbottom}\includegraphics[width=0.75\columnwidth]{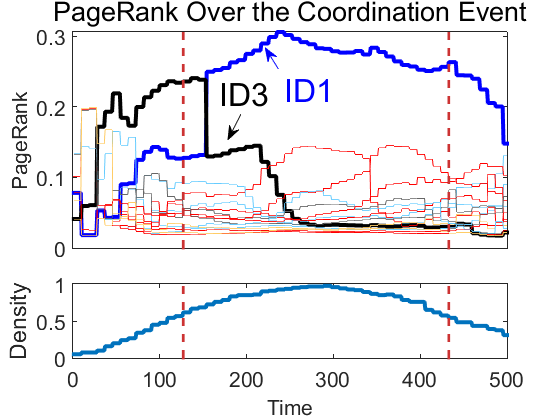}}
\subfigure[t=50]{\label{fig:baboon-exa}\includegraphics[width=0.32\columnwidth]{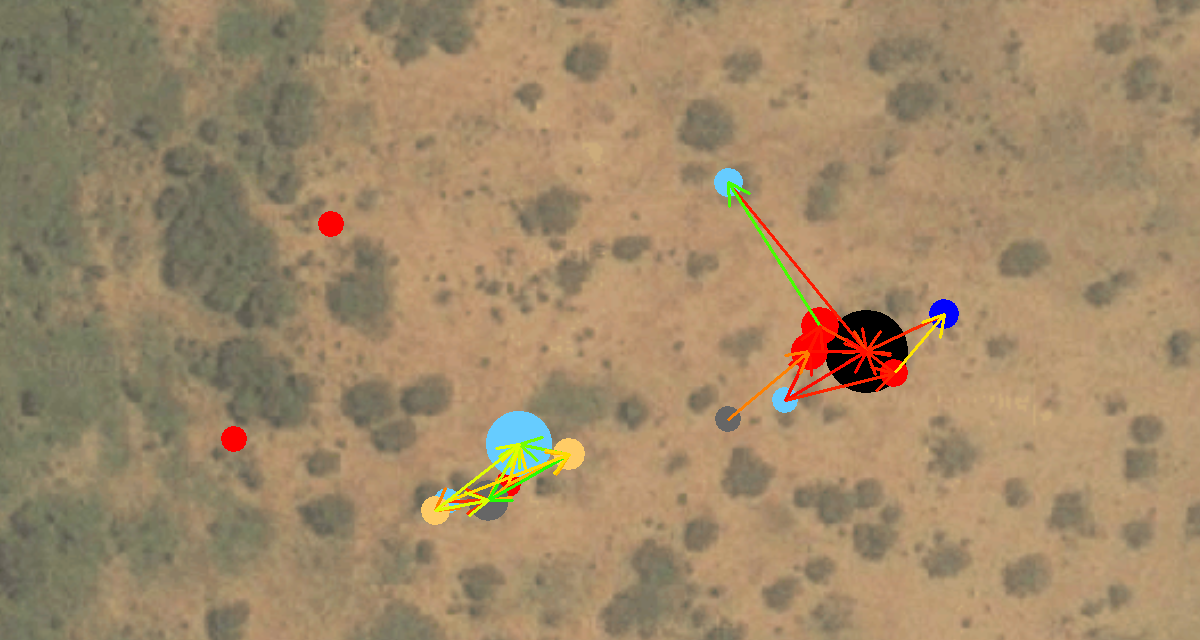}}
\subfigure[t=100]{\label{fig:baboon-exb}\includegraphics[width=0.32\columnwidth]{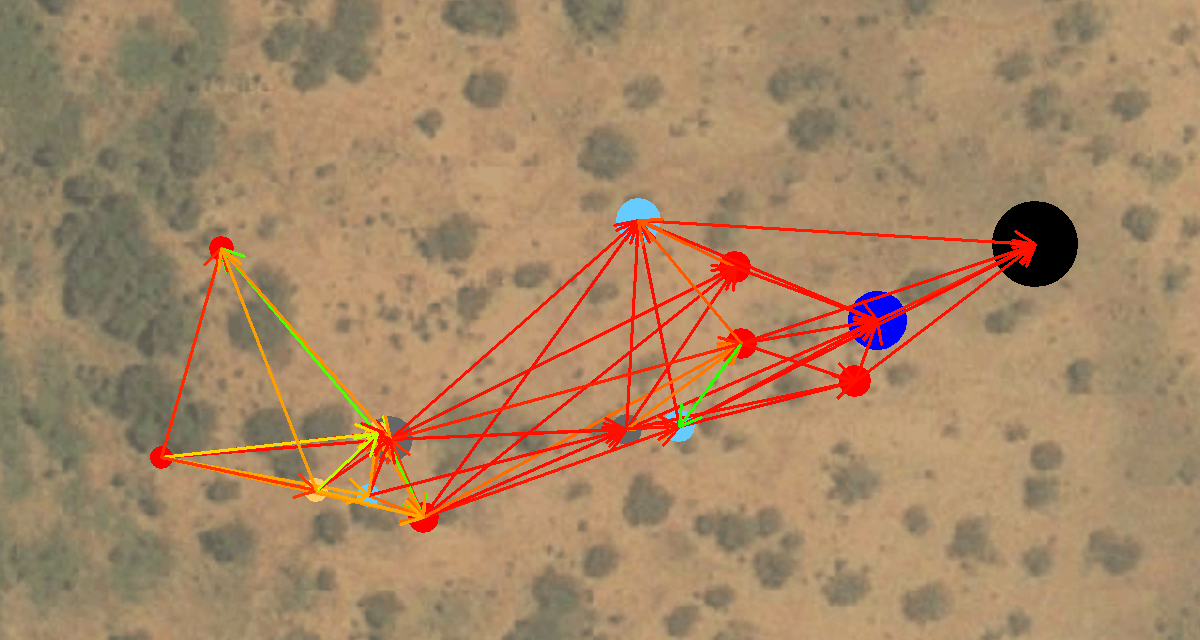}}
\subfigure[t=250]{\label{fig:baboon-exc}\includegraphics[width=0.32\columnwidth]{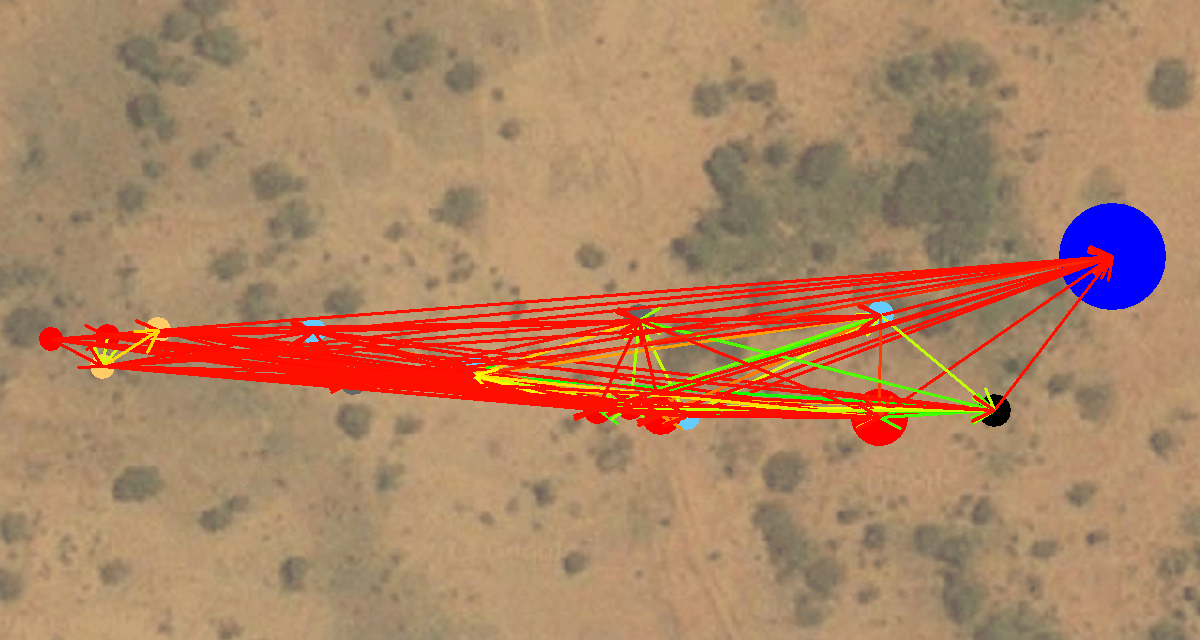}}
%\subfigure{\label{fig:baboon-exB}\includegraphics[width=1\columnwidth]{FIG/BaboonAug12Result-02-02-16-RankOrderPlot}}
\caption{PageRank (top) and density (middle) of the `following' network over time for an event of baboons' movement which initiates by ID3. (Bottom) The locations of individuals over three different time steps ($t=50,100,250$), with the `following' network, and PageRank indicated by node size.}
\label{fig:BaboonRankOrder}
\end{figure}

\subsection{A working example}
\label{subsec:working}
Fig.~\ref{fig:BaboonRankOrder} presents a key example and a brief introduction to our framework, on real GPS trajectory data of olive baboons (\emph{Papio anubis}). Fig.~\ref{fig:baboon-exa}-\ref{fig:baboon-exb} show the leadership of movement of the group by baboon ID3 (Black). Fig.~\ref{fig:baboon-exc} shows the `following' network in the coordination interval. Individual ID3 has the largest PageRank in the first two snapshots but the PageRank of individual ID1 (Blue) surpasses ID3 when the network is `coordinated' (e.g. moving together). If we measure the initiator ranking \emph{after} the network has coalesced, then we miss that ID3 initiated coordination and `built' the network in the pre-coordination interval (to the left of the first dotted red line).\\

We now present each step of the computational framework of FLICA. We will discuss the following relation inference in Section~\ref{subsec:FollNet}, the construction of the following network in Section~\ref{subsec:DyNet}, identification of the coordination interval  and the preceding decision-making period in Section~\ref{sec:CoorInterval}, the identification of the initiator in Section~\ref{sec:RankComp} and the details of model and parameter choices at each step.

\subsection{Following relation inference}
\label{subsec:FollNet}

\begin{figure}[ht!]
\centering
\includegraphics[width=.9\columnwidth]{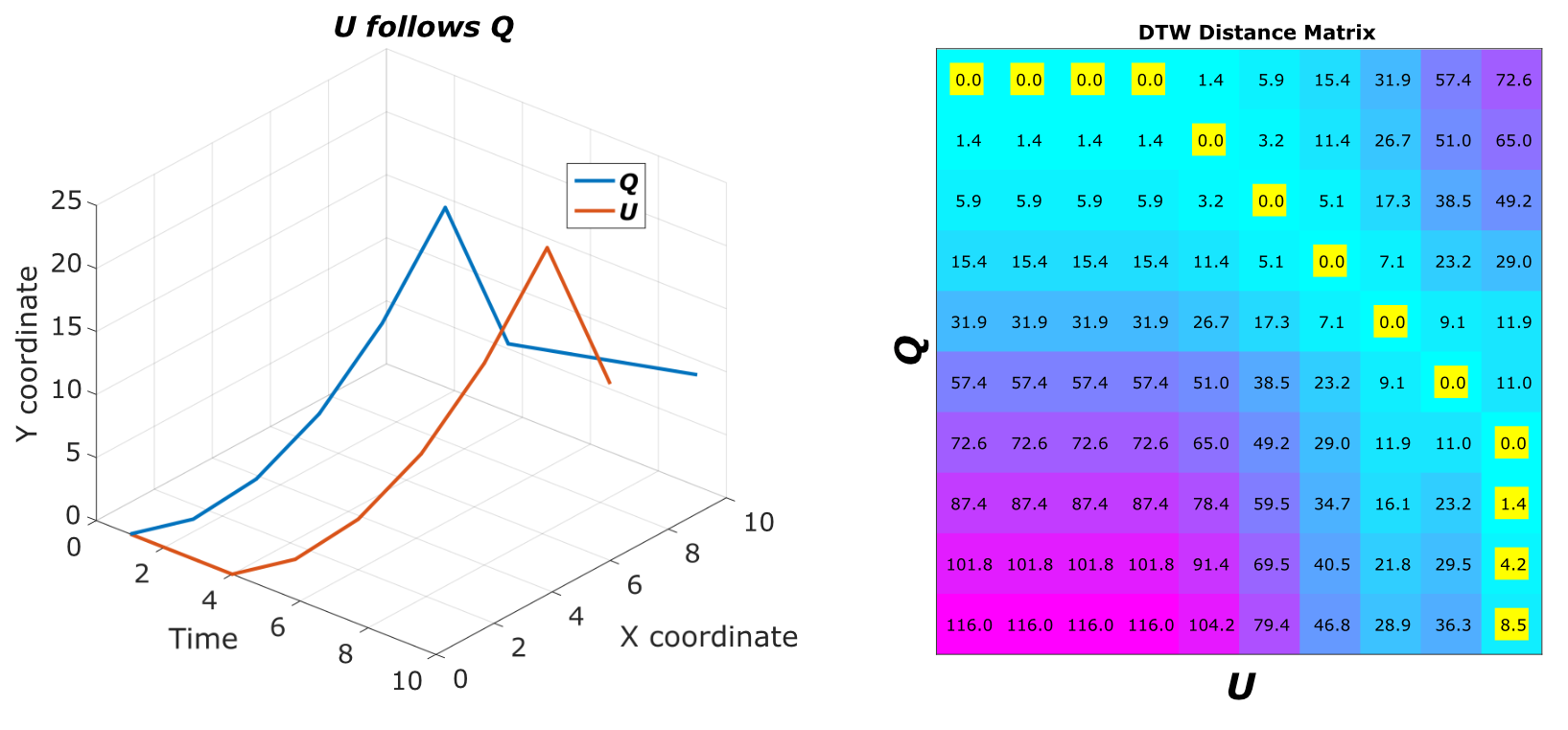}
\caption{(Left) Toy time series showing $U$ following $Q$ with a time delay $\Delta t=3$. (Right) the optimal warping path (yellow boxes) on the $\mathrm{DTW}$ dynamic programming matrix, shifting $U$ backward in time onto $Q$.}
\label{fig:FollReCorr}
\end{figure}

Given a pair of time series $U,Q$, our task here is to find a following relation between $U$ and $Q$. However, we relax a notion of following relation in Def.~\ref{followRDef} to allow some degree of distortion between two time series that follows each other. A measures we need should satisfy two properties. First, a measure must be able to identify common pattern between $U$ and $Q$. A common pattern might not happens in $Q$ the same time as $U$ and the common pattern might have some degree of distortion. Second, a measure must be able to infer a time delay between common patterns in $U$ and $Q$. With these properties, for a $(U, Q)$ pair of time series, we use Dynamic Time Warping ($\mathrm{DTW}$) \cite{Sakoe1978} to measure whether $U$ follows $Q$.  DTW is shown to perform better than several other methods in inferring following relation in time series~\cite{kjargaard2013time} and it is tolerant to noise~\cite{doi:10.1137/1.9781611974010.33}. Fig.~\ref{fig:FollReCorr} (Left) shows two time series, where time-shifting $Q$ ahead in time produces a better match to $U$, illustrated in the \emph{warping path} in Fig.~\ref{fig:FollReCorr} (Right). Let $P_{U,Q}$ be a sequence of index pairs $(i,j)$ which comprise the $\mathrm{DTW}$ optimal warping path of  $(U, Q)$. We compute the mean of the signed index difference over this sequence of index pairs:

\begin{equation}
	\mathrm{s}(P_{U,Q})=\frac{\sum_{(i,j) \in P_{U,Q}}\mathrm{sign}(j-i)}{|P_{U,Q}|}
	\label{eq:traCorr}
\end{equation}
This function measures the extent of warping between two time series. If time series cannot be shifted one-onto-the-other with a consistent positive or negative sign, $|\mathrm{s}(P_{U,Q})| \approx 0$, then there is no following relation between $U$ and $Q$. When $\mathrm{s}(P_{U,Q})$ is positive, $Q$ \emph{follows} $U$, otherwise, $U$ \emph{follows} $Q$. In Fig.~\ref{fig:FollReCorr}, $\mathrm{s}(P_{U,Q}) \approx -3$.
 
\subsection{Dynamic following network inference}
\label{subsec:DyNet}

As shown in Section~\ref{subsec:working}, a coordinated activity is dynamic in the aspect of who leads a group at each time step. Using only summary statistics of static following network to represent the entire coordinated activity cannot capture dynamics of coordinated activity. Therefore, we deploy a dynamic network procedure to analyze coordinated activities in time series, which is a common technique to deal with dynamics of  data \cite{holme2014temporal}.

In our setting, the set of $n$ $m$-multidimensional time series $\mathcal{D}$ ({\em e.g.,} a matrix of size $[n \times m \times t^* ]$), a window size parameter $\omega$, and a window shift parameter $\delta$ (default is $0.1\omega$) are the inputs for our framework. 

\IncMargin{1em}
\begin{alprocedure}[ht!]
\caption{\small CreateDyFollowingNetwork}
\SetKwInOut{Input}{input}\SetKwInOut{Output}{output}
\Input{A set of time series $\mathcal{D}$, a time window $\omega$, and a window shift $\delta$}
\Output{An $n\times n \times t^*$ adjacency matrix $E^*$.}
\BlankLine
{\small
$K \leftarrow (t^* - \omega)/ \delta$ \;
\For{$i\leftarrow 1$ \KwTo $K$}{

\textcolor{blue}{\tcc*[h]{current time interval} } \\ $w(i)=[(i-1)\times \delta,(i-1)\times \delta + \omega]$ \; 
 \textcolor{blue}{\tcc*[h]{SubTimeSeries($\mathcal{D},w(i)$) returns all sub time series in $\mathcal{D}$ within the interval $w(i)$}}
 
$\mathcal{Q}_i \leftarrow$SubTimeSeries($\mathcal{D},w(i)$)\;
$E\leftarrow$CreateFollowingNetwork($\mathcal{Q}_i$) \;

 \textcolor{blue}{\tcc*[h]{Set all edges within the time interval $[(i-1)\times\delta,i\times\delta]$ to be similar}}

$E^*_{ t \in [(i-1)\times\delta,i\times\delta]}\leftarrow E$ \;
}
$\mathcal{Q} \leftarrow$SubTimeSeries($\mathcal{D},[K\times\delta,t^*]$)\;
$E\leftarrow$CreateFollowingNetwork($\mathcal{Q}$) \;
$E^*_{ t \in [K\times\delta,t^*]}\leftarrow E$ \;
}
\label{algo:CreateDyFollowingNetwork3}
\end{alprocedure}\DecMargin{1em} %\todo{Not sure the pseudocode is necessary}

Let the $i$th time interval be given by: $w(i) = [(i-1) \times \delta,(i-1) \times \delta + \omega]$.  For each $w(i)$, we extract a set of sub time series $\mathcal{Q}_i$ from $\mathcal{D}$. The $\mathcal{Q}_i$ is the $[n \times m \times \omega ]$ dimensional matrix of the time series set. Then we construct a following network $G=(V,E)$ as defined in Def.~\ref{followNetDef}. The nodes represent  the time series from $\mathcal{Q}_i$  and $E$ is a set of edges between time series nodes such that if $U,W \in \mathcal{Q}_i$ and $U$ follows $W$ according to Eq.~\ref{eq:traCorr}, then $e_{U,W} \in E$ with the edge weight $|\mathrm{s}(P_{U,W})|$. We calculate a following network for each $w(i)$ to construct a dynamic following network $G^*=(V,E^*)$. The pseudo code is given in Procedure~\ref{algo:CreateDyFollowingNetwork3}.

\subsection{Coordination intervals detection}
\label{sec:CoorInterval}
\begin{figure}[ht!]
\centering
\includegraphics[width=.55\columnwidth]{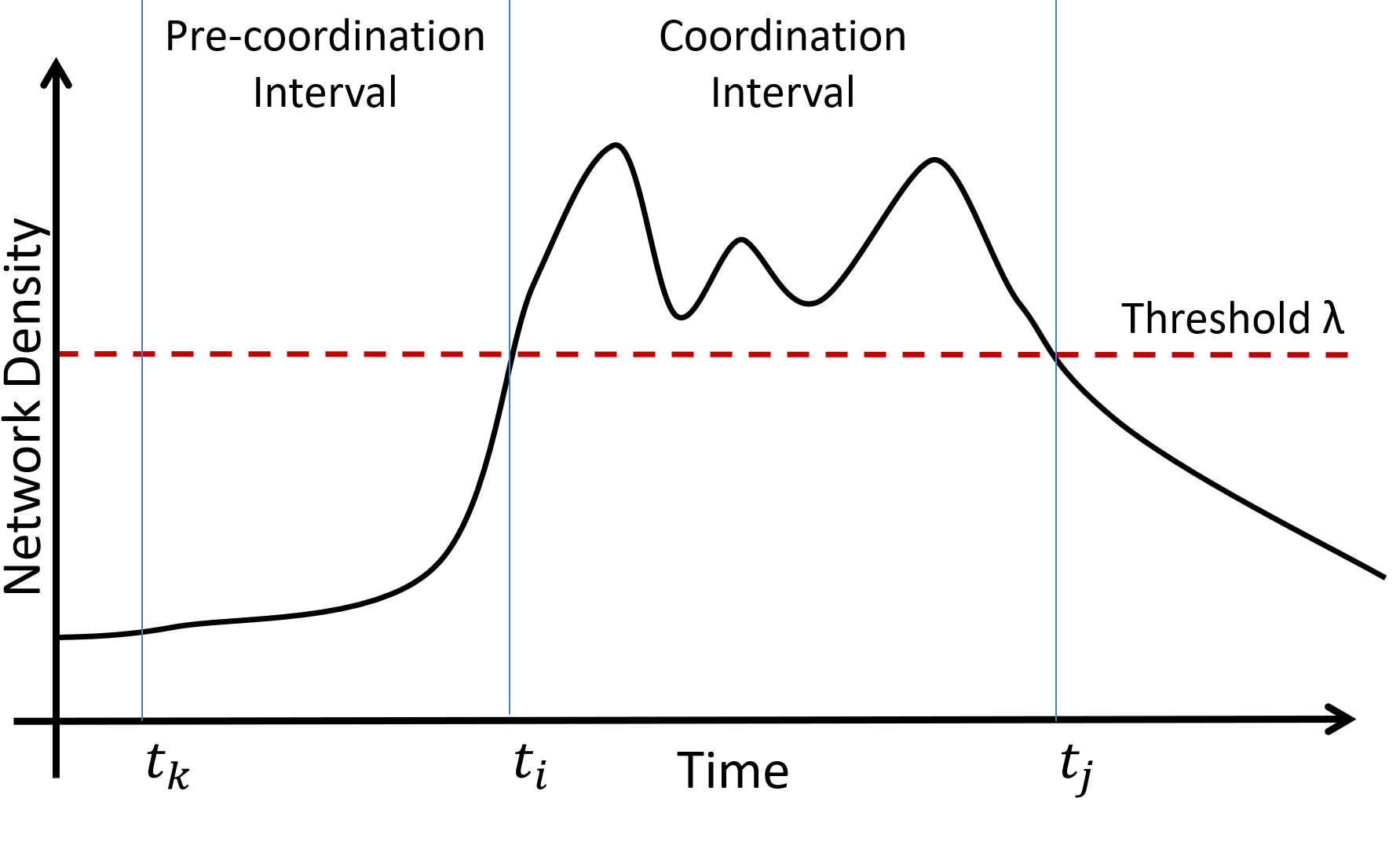}
\caption{A \emph{coordination event} is a pair of intervals. We define the \emph{pre-coordination interval} and \emph{coordination interval} using threshold $\lambda$ on the network density time series.}
\label{fig:CoorEvent}
\end{figure}

\textit{Network density} of the following network serves as the measure of the extent of coordination over all time series pairs (by Def.~\ref{CoorDef}, during the coordination interval {\em every} pair has a following relation.) We can use this observation to identify times of \textit{approximate coordination}.

Given a time series of network densities, denoted by $\mathrm{d}$, over a dynamic following network $G^*$, and a density threshold parameter $\lambda$, the time interval $[t_i,t_j]$ is a $\lambda$-coordination interval if $\mathrm{d}(t) > \lambda$ for \emph{all} $t \in [t_i,t_j]$. The \textit{pre-coordination interval} of coordination $[t_i,t_j]$ is the interval $ [t_k,t_i-1]$, where the discrete derivative $\mathrm{d}(t) -\mathrm{d}(t-1) \geq 0$ for all $t \in [t_k,t_i-1]$. Together, these intervals are one \textit{coordination event}, represented by the 3-tuple of time indices $I=(t_k,t_i,t_j)$. The collection of coordination events is a set $C=\{I_l\}$. All complete event intervals $[t_k, t_j]$ are mutually disjoint in $C$, and $|C|$ denotes the total number of 3-tuples. Fig.~\ref{fig:CoorEvent} illustrates the definition of a \emph{coordination event} as a pair of time intervals. To reduce the number of intervals generated near the threshold $\lambda$, we apply a greedy merging of nearby coordination intervals (taking the range from the window size $\omega$). 

\subsection{Ranking comparison}
\label{sec:RankComp}
%Recall, that by Lemma~\ref{PageRankLemma}, if $U$ follows $V$ then the PageRank of $U$ is less than that of $V$.

On each coordination {event} $I=(t_k,t_i,t_j)$, let $R_I$ be some ranking of individuals within the pre-coordination interval $[t_k,t_i-1]$. We focus on ranking within pre-coordination because this is the interval where coordination is initiated. The {\em global} rank order of pre-coordination, denoted by $\hat{R}$, is the average of all $R_I$ where $I \in C$. 

We measure {\bf initiator ranking} according to three different methods: PageRank~\cite{PageRank_Brin1998107}, velocity convex hull (VCH), and position convex hull (PCH). Recall, that by Lemma~\ref{PageRankLemma}, if $U$ follows $V$, then the PageRank of $U$ is less than that of $V$. Thus, the initiator is expected to have the highest PageRank. VCH measures how often an individual moves faster than others. It represents a model of leadership for movement. This model can be found in many social species \cite{Dyer:2009aa,Stueckle2008}. PCH measures how often an individual moves to an area before others. For example, in a flock model~\cite{andersson2008reporting}, a leader is positioned at the front of the group's trajectory.

\subsubsection{PageRank}
PageRank is a standard method for measuring the importance of a node recursively by the importance of the nodes linking to it. In a directed network where a link represents a \emph{following} relations between nodes, PageRank measures `following' \emph{paths} passing through a particular node. Thus, it fits well with our definition of leadership.
%and can be thought of as approximating the proportion of time spent at a node over many random walks on the network. 

PageRank returns a weight vector of length $n$, with a sum of 1. For each time step $t$, we calculate PageRank for each static graph $G_t$ within a dynamic following network $G^*=(V,E^*)$. Let $\mathcal{R}=(R_{pr, 1},\dots, R_{pr,t^*})$ be a sequence of $n$-length PageRank Order vectors where $R_{pr, t} = \mathrm{argsort}(PageRank(G_t))$ such that $R(i)_{pr, t}$ represents the rank of individual $i$ and $R(i)_{pr, t}<R(j)_{pr, t}$ if the PageRank value in Eq.~\ref{eq:PReq} of $i$ is greater than the value of j ( $\pi_i>\pi_j$.)  The leader $L$ at time $t$ is the individual who has the highest value of PageRank $\pi_L$ or $R(L)_{pr, t}=1$ .  Note that $\mathrm{argsort}(\bullet)$ returns the index list of sorted values w.r.t. descending order.

\subsubsection{Velocity Convex Hull}
In the next two sections, the $s$-energy work by Chazelle~\cite{doi:10.1137/100791671} motivates the use of the convex hull as the measure of the level of initiation of a state change for the group. Chazelle showed that if every agent in a group remains within the convex hull of its neighbors (even if the neighbors change) at each time step, then the system converges to an equilibrium. Thus, to change the steady state, somebody needs to break out of the convex hull of their neighbors. In the initial state, all individuals states such as velocity or position are inside the group's convex hulls of that state. Then after the group decides to change its state, some individuals must step outside the group convex hull to make the change. Hence, by using convex hull analysis, we can measure whether the initiators are also state changers. Specifically, we use convex hull (of position and velocity) analysis to characterize leadership models.\\

The velocity convex hull measures the frequency with which the discrete time series derivative ($dQ/dt$) associated with a node $i$ is outside the bounds of the population's discrete derivative distribution  (including node $i$) in the previous time step. In aggregate, a high rank of this measure indicates which node first `moves' in the group. 

The convex hull can be computed on arbitrary $m$ dimensions of a multidimensional time series, or their derivatives, jointly or independently. The convex hull function $ \mathrm{CH}(\bullet)$ returns an $m$-dimensional surface represented as lines between points in the input data, which encompass all other points. 

Let $\mathbb{V}$ be a $[n \times t-1]$-sized matrix measuring individual velocity over time, on time series dataset $\mathcal{D}$, which is a $[n \times m \times t^* ]$-sized matrix . For an individual $i$ at time-step $t$, we define the following indicator function: \footnote{ We use `*' subscript notation in matrices to indicate slicing in the dimension(s).}
\begin{equation}
  \mathrm{VCH}(\mathbb{V}, i, t)=\left\{
  \begin{array}{@{}ll@{}}
	1, & \mathbb{V}_{i,t} > \mathrm{max}(\mathbb{V}_{*,t-1}) \\
   -1, & \mathbb{V}_{i,t} < \mathrm{min}(\mathbb{V}_{*,t-1}) \\
    0, & \text{otherwise}
  \end{array}\right.
\label{eq:velConvx}
\end{equation}
For time step $j$ we output an $n$-length rank order vector as $R_{v, t} = \mathrm{argsort}((\mathrm{VCH}(\mathbb{V}, i, t))_{i=1...n})$.

\subsubsection{Position Convex Hull}
The position convex hull is analogous to velocity, except that our indicator function measures an individual's position relative to the convex hull containing the population at the previous time step. Rather than look at velocity of initiation, this measure captures an individual's frequency of moving outside the geometric boundaries of the group in the time series space, and close to the average heading of the group (e.g. in `front' of the group).  

We compute the convex hull function on time-step $t$, $H_{t}$ = $\mathrm{CH}(\mathcal{D}_{*, *, t})$, and also introduce the heading vector of individual $i$: $\vec{v}_{i,t} = (\mathcal{D}_{i,*,t-1}, D_{i,*,t})$, and the population heading vector: $\vec{v}_{t}$ = $(1/n)\sum_{i=1..n} \vec{v}_{i,t}$. We define the function $\mathrm{IN}(\text{A, B})$ to denote standard `B contains A' spatial queries between two geometry objects, and $\measuredangle(\vec{v_1}, \vec{v_2})$ to measure the angle between two vectors $\vec{v_1}$ and $\vec{v_2}$.

Using these definitions, we define the position convex hull indicator function for individual $i$ at time $t$:

\begin{equation}
  \mathrm{PCH}(D,i,t)=
  \begin{cases*}	
	1, & $\neg\mathrm{IN}(\mathcal{D}_{i,*,t},H_{t-1}), \measuredangle(\vec{v}_{i,t}, \vec{v}_{t}) \leq 90^{\circ}$\\
   -1, & $\neg\mathrm{IN}(\mathcal{D}_{i,*,t},H_{t-1}), \measuredangle(\vec{v}_{i,t}, \vec{v}_{t}) > 90^{\circ}$\\
    0, & otherwise
  \end{cases*}
\label{eq:posConvx}
\end{equation}

For time step $t$ we output an $n$-length rank order vector as $R_{p, t} = \mathrm{argsort}((\mathrm{PCH}(D, i, t))_{i=1...n})$. %Like the 

\subsection{Leadership model features}
\label{subsec:features-rank}
% For all $I \in C$, and node $i$, \emph{Leadership Support} is defined relative to a particular measure (e.g. PageRank) for $i$ as the fraction of intervals where $i$ is first ranked: $\mathrm{sup_{\bullet}}(i)= \frac{\#((i, 1) \in R_{\bullet, I} )}{|C|},\text{   } I \in C$.
Let the global rank ordering of pre-coordination for PageRank be denoted by $\hat{R}_{pr}$, for VCH by $\hat{R}_{v}$, and for PCH by $\hat{R}_{p}$.  To measure global leadership of pre-coordination in our framework, we order individual nodes $i$ based on initiation support with respect to one of these ranking methods, $R_{\bullet, I}$ , over all coordination events $I \in C$. For example, $R_{pr, I}$ is PageRank-rank-ordered list at the coordination event $I$. If an individual $i$ is at 1st rank at $I$, then $(i, 1) \in R_{pr, I}$; $i$ is an initiator. The initiation support for a node $i$ is the fraction of coordination events at which it was ranked 1 (by a ranking measure $R_{\bullet}$):
\begin{equation}
\mathrm{sup_{\bullet}}(i)= \frac{\left |{\{\forall I\in C \: (i, 1) \in R_{\bullet, I}\}}\right | }{|C|}
\label{eq:LeaderSup}
\end{equation}

We use the Kendall rank correlation coefficient $\tau()$ ~\cite{Kendall1938} to compare event-local and global rank-orders. %This measure provides a similarity between two rankings according to their ordinal agreement over all list-pairs (e.g. $j$ is ``below'' $i$ in both lists). 
To compare global and local rank orders, we use the mean Kendall rank correlation over all coordination events against the global by Eq.~\ref{eq:LeaderglobalCorr}. For example, $\mathrm{corr}_{v}$ compares local and global velocity-convex-hull-rank orders. 

\begin{equation}
\mathrm{corr}_{\bullet}=\frac{\sum_{I \in C} \tau(\hat{R}_{\bullet}, R_{\bullet, I})}{|C|}
\label{eq:LeaderglobalCorr}
\end{equation}

Similarly, we compute the mean Kendall correlation between local rankings associated with different measures (e.g. VCH, PCH) by Eq.~\ref{eq:LeaderPairwiseCorr}.
%Similarly, we report cross-domain mean of Kendall rank correlation over local rankings:

\begin{equation}
\mathrm{corr}_{\bullet, \bullet}=\frac{\sum_{I \in C} \tau(R_{\bullet, I}, R_{\bullet, I})}{|C|}
\label{eq:LeaderPairwiseCorr}
\end{equation}

$\mathrm{corr}_{\bullet}$ formalizes our intuition that leaders consistently move outside of the spatial extent ($\mathrm{corr}_{p}$), or the distribution of velocity over the population ($\mathrm{corr}_{v}$). By comparing the global vs. local correlation in rank ordering, we measure the stability of the global ranking is over time. %A low correlation suggests that the hidden distribution of real leadership may have changed over the length of the input.   

$\mathrm{corr}_{\bullet, \bullet}$ measures the relationship between higher-order graph structure and simple time series features. Using this measure, we can gain a better understanding of the high-level aspects of initiating coordination. For example, we see whether changing velocity ($\mathrm{corr}_{v, pr}$), or position ($\mathrm{corr}_{p, pr}$) within the group is correlated with network rank position.

\subsection{Local vs. Global Matching}
\label{subsubsec:local}

\begin{figure}
\centering     %%% not \center
%\captionsetup[subfigure]{labelformat=empty}
\subfigure[Pattern order alignment]{\label{fig:sparse-exA}\includegraphics[width=.48\columnwidth]{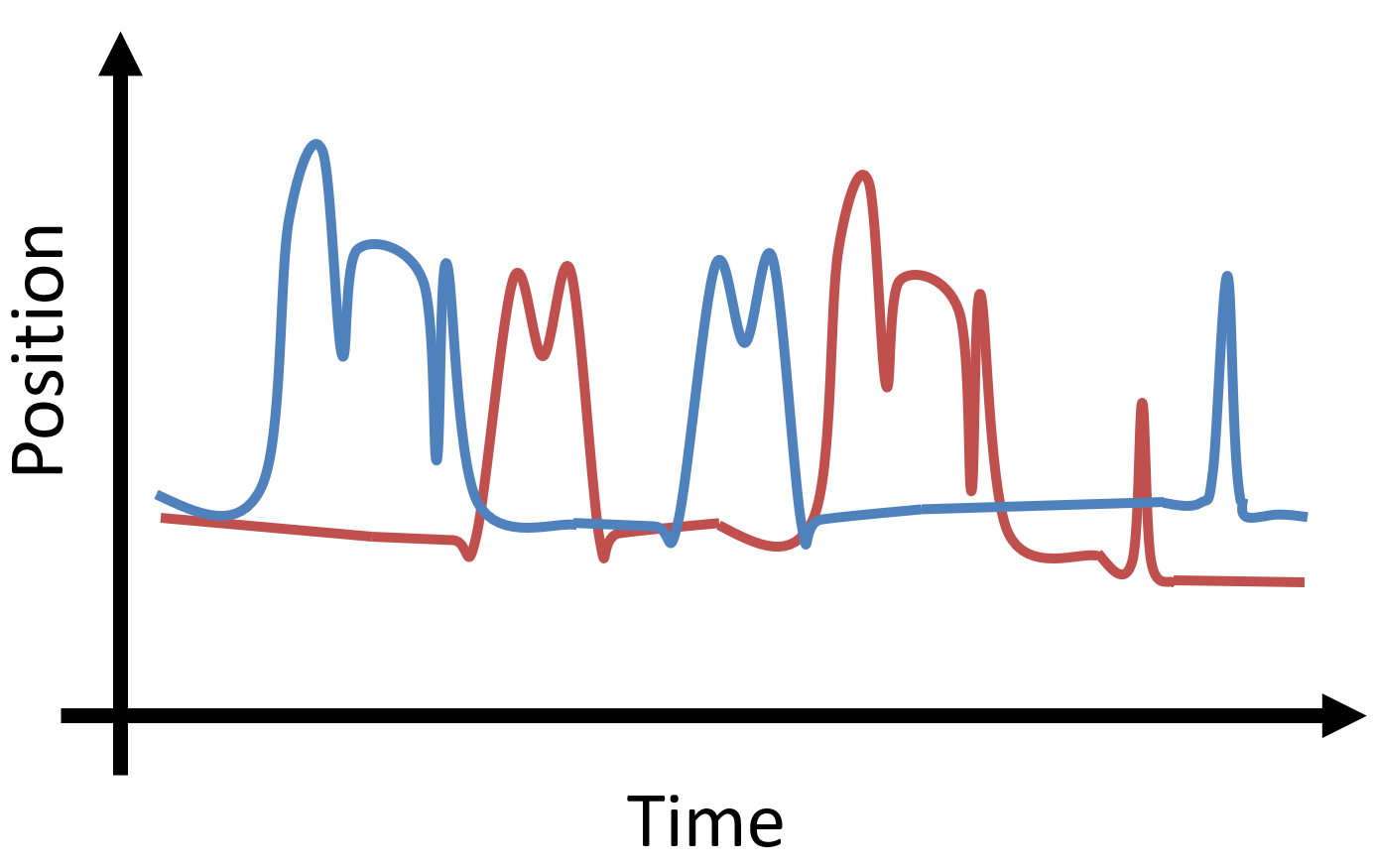}}
\subfigure[Global shift alignment]{\label{fig:sparse-exB}\includegraphics[width=.48\columnwidth]{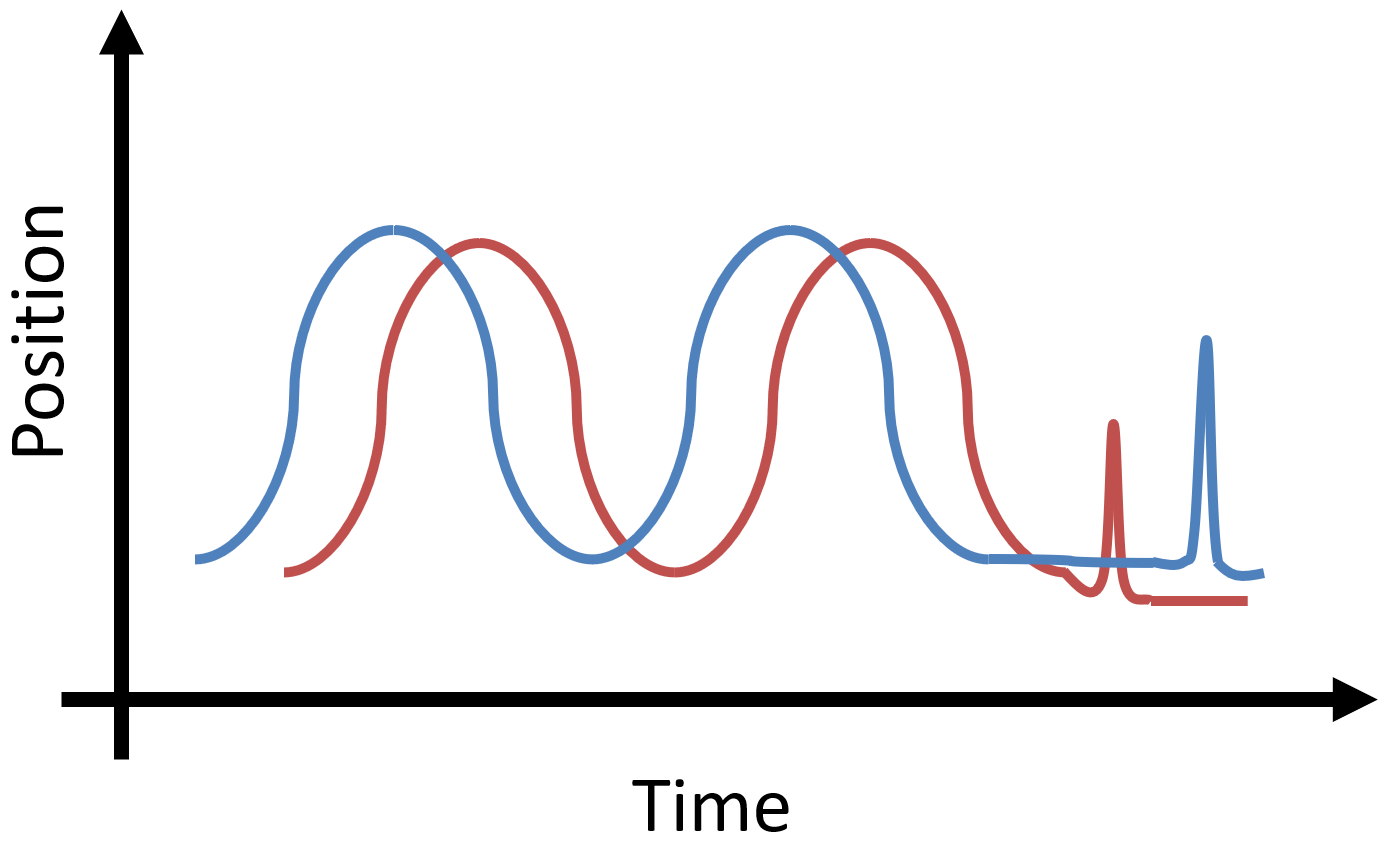}}
\caption{Dynamic Time Warping global vs. local example}
\label{fig:globalvlocal}
\end{figure}

Our proposed framework uses local alignment on time series subsequences, rather than global alignment on the full time series.  Fig.~\ref{fig:globalvlocal} presents a motivation for this choice. Suppose we intend to match \emph{sparse} `following' events represented as the pair of spikes with relatively low magnitude at the end of the red and blue time series. In Fig.~\ref{fig:sparse-exA}, the time series is shifted to match one of the two patterns, depending on the cost. This forces a mismatch of the `following' event. Similarly, Fig.~\ref{fig:sparse-exB} has a low cost matching by shifting the entire time series at a constant rate. By matching only local subsequences, we can recover both of these `following' events.

\section{Experimental setup}
We evaluate our framework on eight synthetic movement trajectory models and three real datasets. 

%In this section,  we consider details regarding description of each dataset we use to obtain framework results in both synthetic and biological datasets. We begin with simulation models to find out that the framework actually works when we have a knowledge of truth about a leader. Afterward, we concentrate on real biological data to see the effects of framework. 

\subsection{Simulation models}

%We develop synthetic trajectory models which capture several hypotheses of leadership within time series dynamics. This inexhaustive collection are not intended to represent individual movement with high fidelity, but to provide contrasting aspects for model selection and comparison. We propose eight synthetic models: dictatorship, hierarchical, linear threshold, independent cascade, crowd, initiator, event-based, and fixed-destination (random) models. For evaluation, we attempt to identify the top-ranked individual associated with the ground-truth label in the simulation.

%control in the simulation We are primarily interested in identifying the top-ranked leader under different models  
%Examining Using simulations , and using simulations

\subsubsection{Dictatorship model (DM)}
\begin{figure}[ht!]
\centering
\includegraphics[width=0.60\columnwidth]{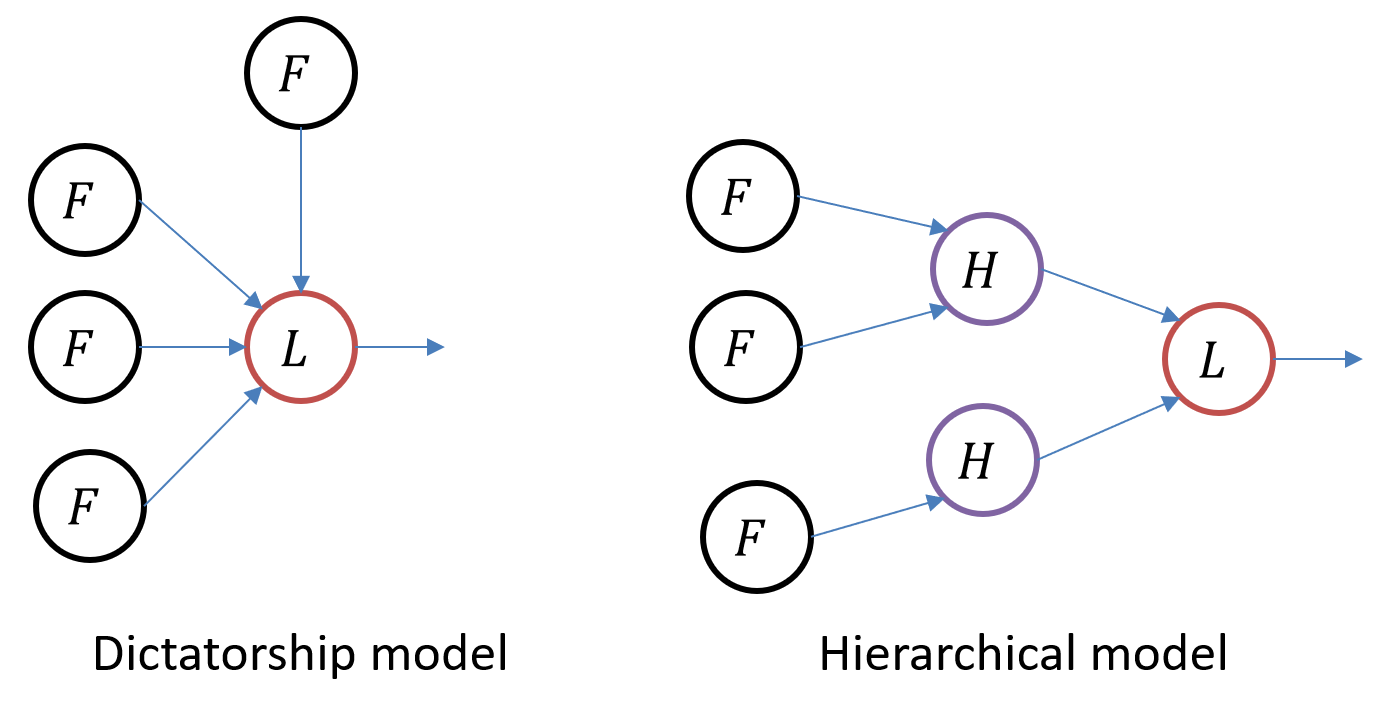}
\caption{Examples of individual movements from Dictatorship and Hierarchical model. Nodes represent individual positions and arrows represent directions of individual's movement. (Left) in Dictatorship model, everyone follows a leader $L$, while there is a hierarchy to follow for each individuals in Hierarchical model (right). }
\label{fig:DMHMpic}
\end{figure}

\begin{figure}[ht!]
\centering
\includegraphics[width=0.7\columnwidth]{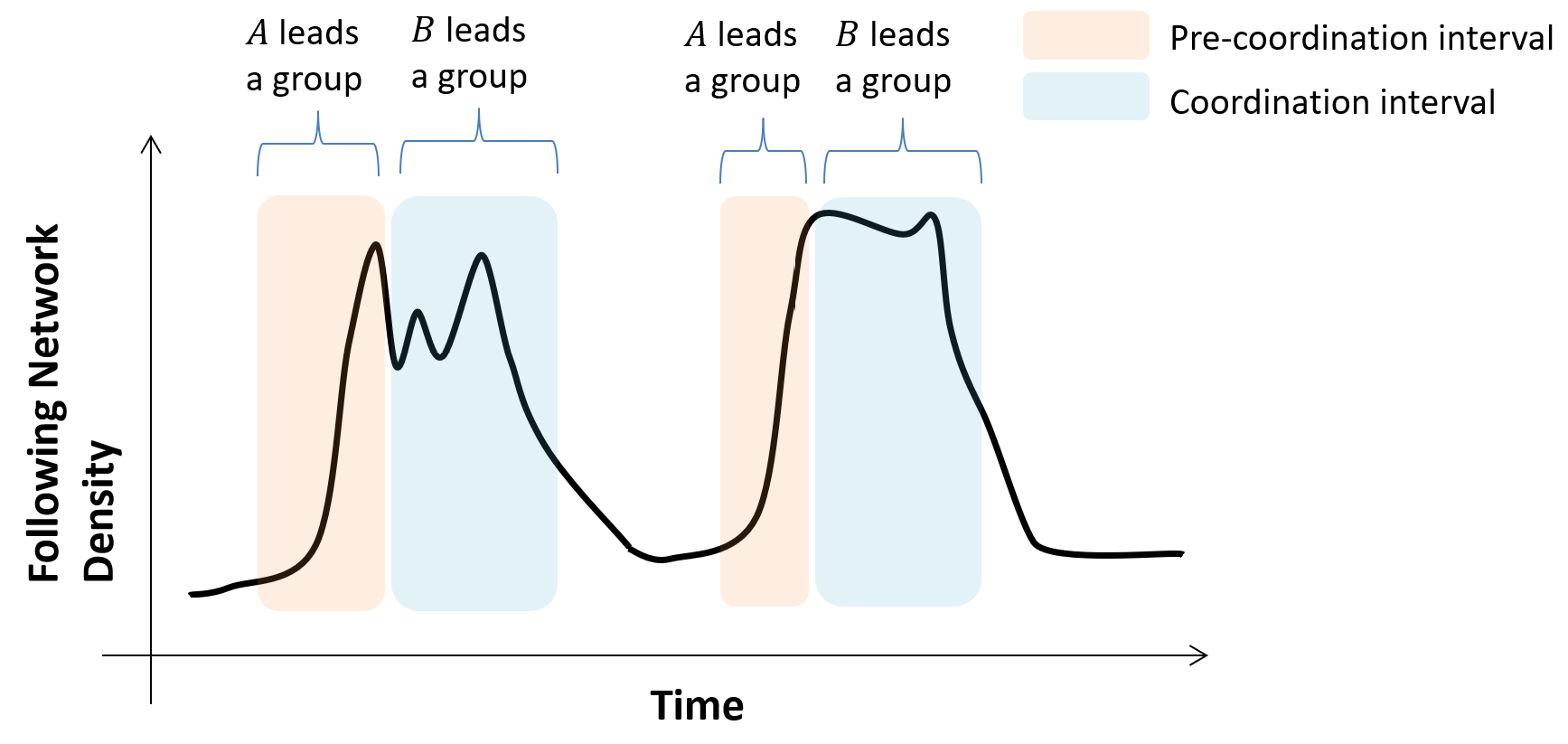}
\caption{An example of following-network-density time series of Switching Dictatorship model . There are two coordination events where individual $A$ leads a group in both pre-coordination intervals while $B$ leads a group during coordination intervals.}
\label{fig:InitDypic}
\end{figure}

In this model, we fix a single initiator who initiates movement from initial positions of the population. At the start of the pre-coordination interval, the initiator moves in a fixed direction and acceleration. Other individuals wait for a randomly sampled lag, before following the initiator at a fixed acceleration (with sampled noise in the heading). After a fixed duration of coordinated movement over the entire population, individuals decelerate at random, until stopping. Fig.~\ref{fig:DMHMpic} (left) shows the example of DM.  The Switching Dictatorship model (DM-S) selects two fixed individuals over each trial: a single individual as an initiator during pre-coordination, and another single individual as `initiator' during coordination. Fig.~\ref{fig:InitDypic} shows an example of following-network-density time series of DM-S. 

\subsubsection{Hierarchical model (HM)} %*********************
This model is a variation of DM, where we fix a number of individuals (n=4) to follow the previous individual in the sequence, after a sampled lag. The remainder of individuals in the population follow exactly one of these high-ranking individuals, allocated in decreasing proportion per rank.  Fig.~\ref{fig:DMHMpic} (right) shows the example of HM. The Switching Hierarchical model (HM-S), similarly to DM-S, selects unique pairs of individuals for each hierarchy level, switching after the pre-coordination interval as in DM-S.

\begin{figure}[ht!]
\centering
\includegraphics[width=0.7\columnwidth]{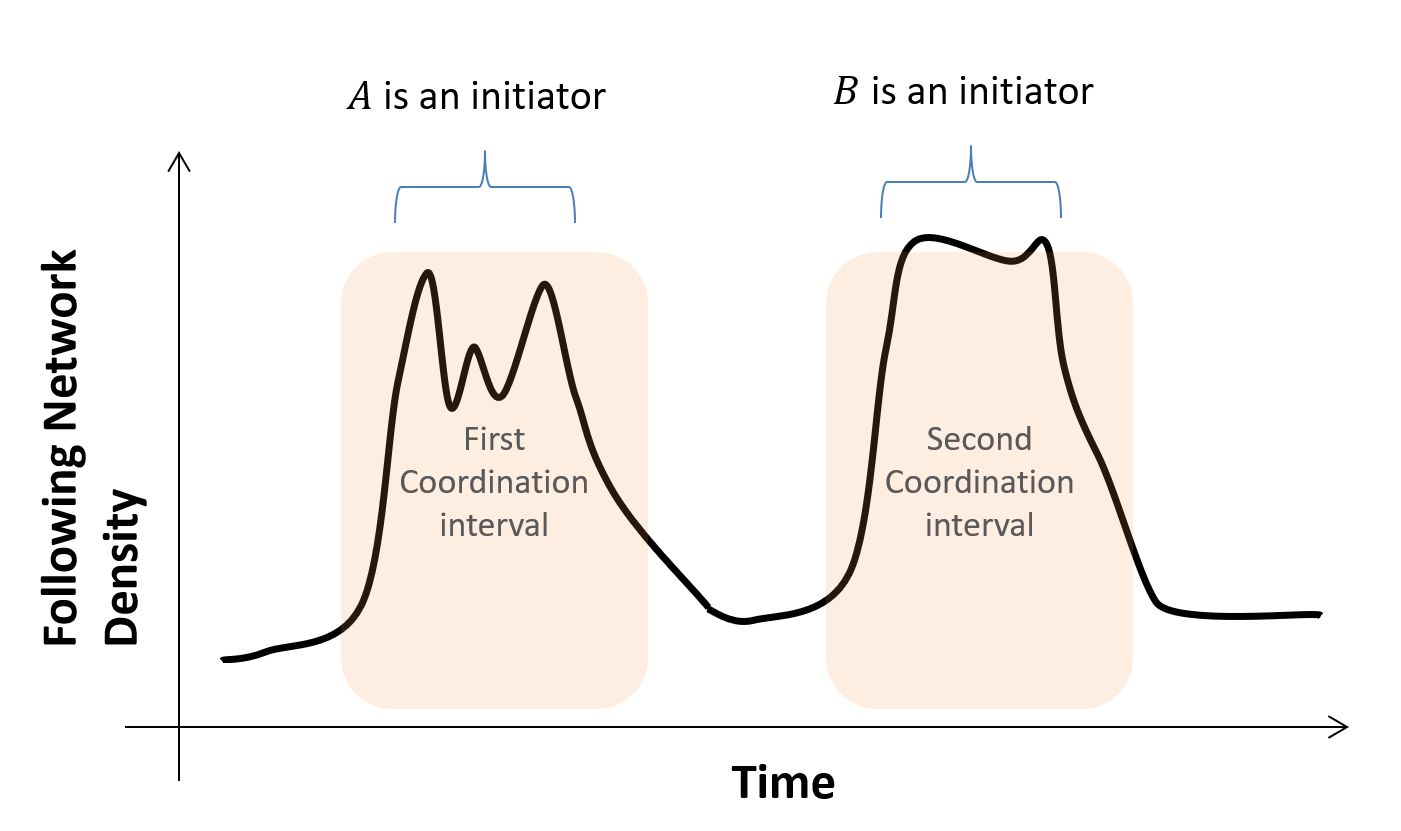}
\caption{An example of following-network-density time series of Event-based model. There are two coordination events where individual $A$ leads a group in the first coordination event while $B$ leads a group during the second coordination event.}
\label{fig:EMpic}
\end{figure}

\subsubsection{Event-based model (EM)}
This model is a variation of the Dictatorship model where each coordination event has a different, unique initiator. For example, in one of our applications, a troop of baboons may follow an initiator to a food source in the morning, and follow a different initiator in the evening to the sleeping site. No existing methods can infer these two situations except our framework.  Fig.~\ref{fig:EMpic} shows an example of following-network-density time series of EM. 

\begin{figure}[ht!]
\centering
\includegraphics[width=0.7\columnwidth]{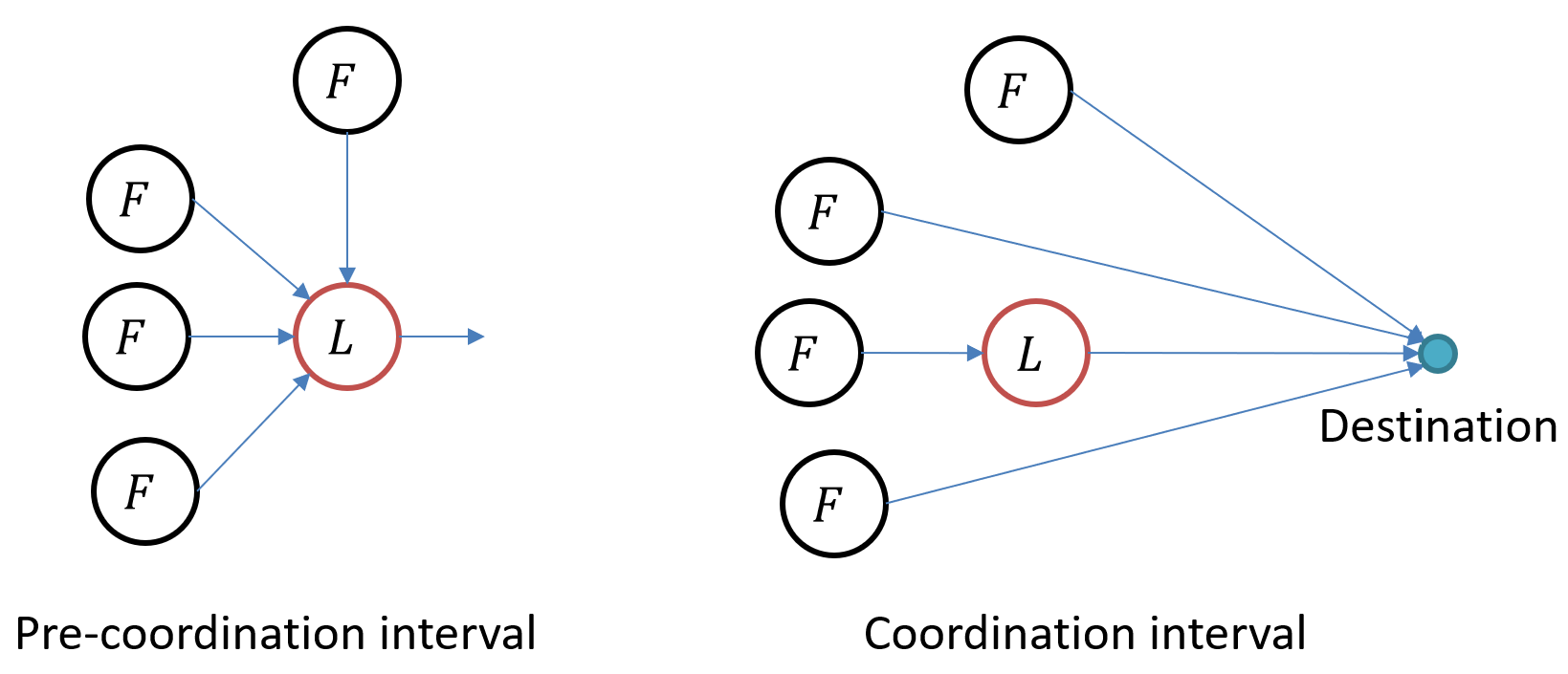}
\caption{Examples of individual movements from  Initiator model. Nodes represent individual positions and arrows represent directions of individual's movement. (Left) during a pre-coordination interval, everyone follows a leader $L$, while, during a coordination interval, everyone knows the direction and moves to the destination directly without following a leader (right). }
\label{fig:Initpic}
\end{figure}

\subsubsection{Initiator model (INIT-k)}
In this model, we fix $k$ initiators who initiate movement from random initial positions of the population. At the start of the pre-coordination interval, all initiators move on a single target. Non-initiators move in randomly sampled directions with a fix velocity, then follow their initiators after a random time lag. After the pre-coordination period, all individuals move toward a single target, without following their initiators. The example of INIT-k model is at Fig.~\ref{fig:Initpic}. We run simulations for INIT-1 and INIT-4 initiator models.

  %The Multiple Initiators Model (`INIT-M'),  which is similar to InitM except there are four leaders.

%\ivan{``each leader moves in a fixed direction and velocity'' are these to the same destination? random location?}
%\ivan{this model is from a citation, right?}
%\ivan{I need this clarified. Also idea: could this be parameterized on the height of the tree and the max number of children? -- followers always follow their leader.}

%For the remainder   of individuals who follow each other after a fixed time where a sample  can be viewed as a special case of previous sole dictatorship model but it has a bit more restrict constraint; a leader start moving first, then high-rank individuals, and ordinary members move lastly. Formally, a leader initiate moving in the beginning of pre-coordination, then high-rank members start following their leader after some delay time. Finally, others start following either/both high rank individual or/and their leader. In our simulation, there are three high-rank members following the leader before low-rank members. Hence, for this model, the $Sup, Corr_{vel}, Corr_{pos}$, and $Corr_{pr,vel}$ features should be high, but others can be vary the same as previous one. 

\begin{figure}[ht!]
\centering
\includegraphics[width=0.7\columnwidth]{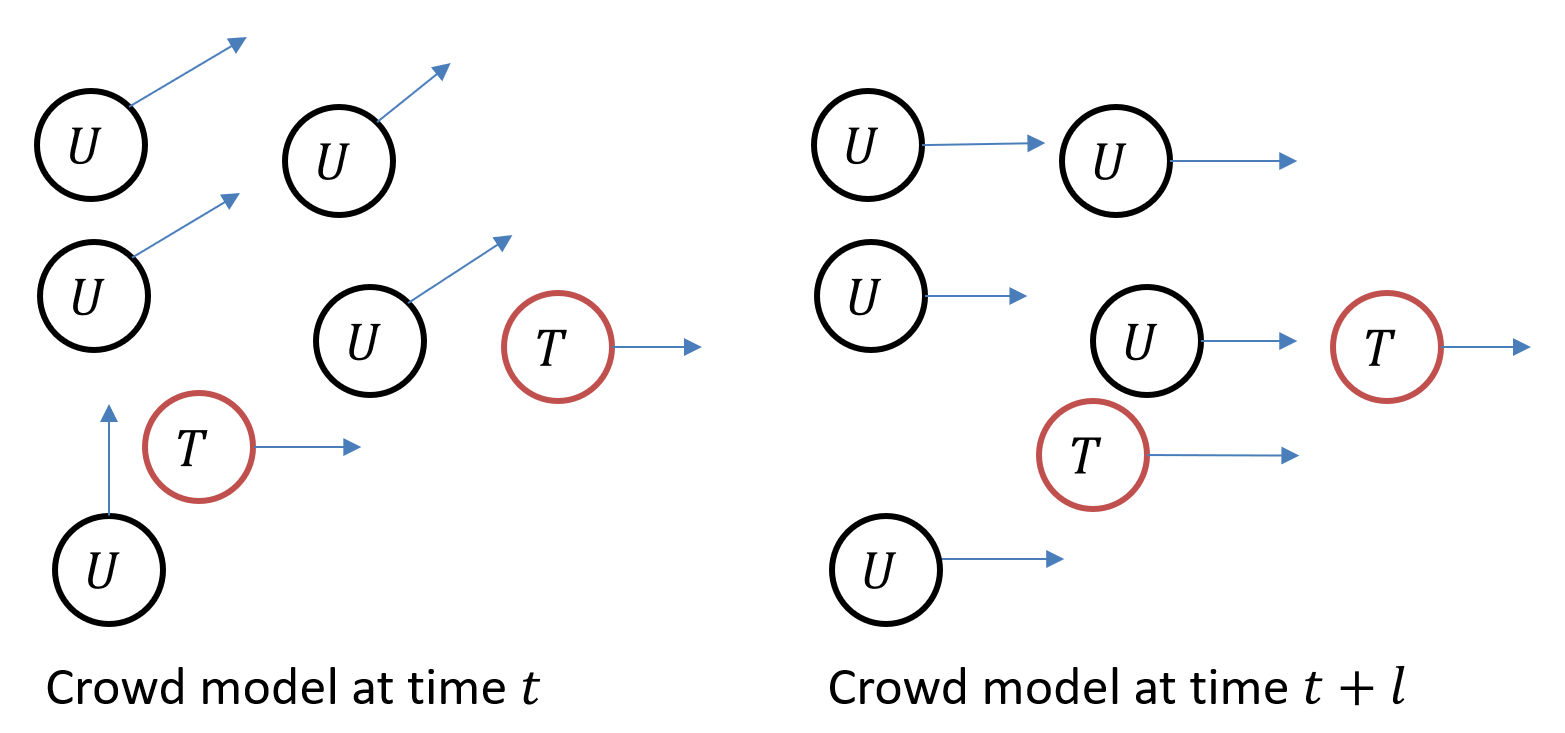}
\caption{Examples of individual movements from  Crowd model. Nodes represent individual positions and arrows represent directions of individual's movement. (Left) at time $t$, everyone follows some directions except informed individuals ($T$ nodes) which moves directly to a target. Then, at time $t+l$ (right), the group's direction, which is the average of individual's directions, gradually changes toward the target. }
\label{fig:CMpic}
\end{figure}

\subsubsection{Crowd model (CM)}
This model~\cite{Song2014crowdmodel} is a collective movement model where $k$ (=4) informed individuals move toward a target, and the remaining (=16) uninformed individuals move in a linear combination of a direction toward the group's centroid, and the average direction of the group. The example of CM is at Fig.~\ref{fig:CMpic}.

\begin{figure}[ht!]
\centering
\includegraphics[width=0.7\columnwidth]{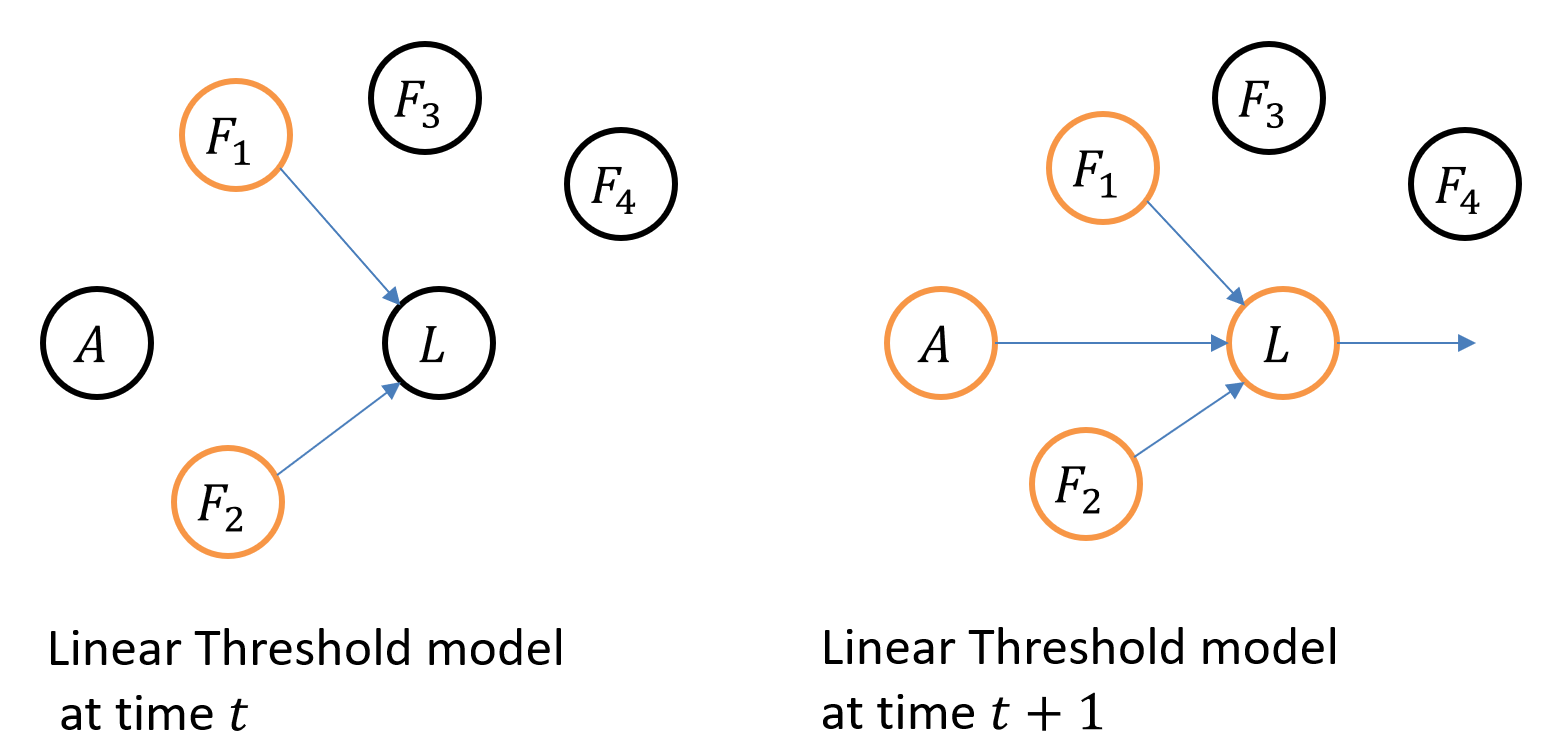}
\caption{Examples of individual movements from  Linear Threshold model. Nodes represent individual positions and arrows represent directions of individual's movement. (Left) at time $t$, only active individuals (orange color) move toward an initiator $L$. Suppose $k=3$ and $\rho=0.50$, at time $t$, an inactive individual $A$ has two active individuals $F_1,F_2$ and $L$ as its neighbors. Since 66\% of $A$'s neighbors are activated, then, at time $t+1$ (right), $A$ is active and start moving toward $L$. }
\label{fig:LTpic}
\end{figure}

\subsubsection{Linear Threshold model (LT)}
This model~\cite{kempe2003maximizing} initiates individual movement by propagation of a linear threshold process on the dynamic network, defined by the $k$-nearest neighbors at the current time-step. The model is parameterized by $\rho$, the proportion of these $k$ neighbors required to be infected in order to initiate movement. Once activated, the individual follows a single initiator. The initial probability of activation for each individual is $0.5$. We explore the parameter space on combinations of: $k \in \{3,5,10\}$ and $\rho \in \{0.25,0.50,0.75\}$. The example of LT is at Fig.~\ref{fig:LTpic}.

\subsubsection{Independent Cascade model (IC)}
This model~\cite{kempe2003maximizing} is another propagation process similar to LT. At each time step, each active individual moves toward the initiator and independently attempts to activate its $k$-nearest neighbors with the probability of $\rho$. If the individual fails to activate a neighbor, it cannot attempt to activate the same neighbor again. We explore the same sample parameter space as in the LT model.

\subsubsection{Random model}
In this model, there is no `following' relations. At the start of the pre-coordination interval, all individuals start moving to a fixed direction, independently of others in the population. We expect the relative positions of individuals to yield some following relations only by chance. 

%This model is partially like SD that everyone moves together to specific destinations except there is no leader. Basically, $Corr_{vel}$ and $Corr_{pos}$ can be high since there might be someone always explore new areas and/or moving start moving before others.  Since it is a random model, therefore, no one follow these individuals and $Sup, Corr_{pr,vel}$ and $Corr_{pr,pos}$ should be close to zero.

%\ivan{Do they all start at the same time? Isn't it too trivial? What is the following network density?}

%\ivan{Do individuals sample different fixed velocities? How: ``$Corr_{vel}$ and $Corr_{pos}$ can be high since there might be someone always explore new areas and/or moving start moving before others''}

%\ivan{what about assigning the leader label as the individual who arrives at the destination first? (or more generally, the leader ranking by order of arrival) this actually tests a claim that the model doesn't only find leaders at the head of the group.}

\subsection{Synthetic trajectory simulation}
For each of the above models, we generate a trial of synthetic data consisting of $20$ individuals, and $20$ separate coordination events, for a total of 12,000 time-steps. Each coordination event has pre-coordination and coordination intervals of $200$ time-steps each. Following the coordination interval is another $200$ time steps of a post-coordination before repeating. We generate $100$ trials for each of models. In total, we have 2,700 simulation datasets.

%\ivan{18,000? 3*300*20}

%of $300$ time-steps, a coordination interval of 300  There are $20$ coordination intervals For all simulation data, they consist of 20 individuals trajectories of which their length are 12,000 time steps, and there are 20 cycles of coordination for each dataset. In each cycle, it can be divided into 300 time steps for pre-coordination, 300 for during coordination, and 300 time step post-coordination. After the end of each cycle, the next cycle continue again at pre-coordination.  

\subsection{Real datasets}
%We demonstrate the utility of our framework on three real-world datasets from two different domains. First, we analyze biological trajectory datasets derived from GPS, and cameras. Next we look at fifteen years of stock closing-price data from the NASDAQ index.  

%In order to evaluate the performance of our framework, we deploy two biological datasets: baboon trajectories and fish trajectories.

\subsubsection{Baboon trajectories}
\label{subsubsec:baboon}
High-resolution GPS collars track 26 individuals of a troop of olive baboons (\emph{Papio anubis}) living in the wild in Mpala Research Centre, Kenya \cite{crofoot2015data,Strandburg-Peshkin1358}. The data consists of latitude-longitude location pairs for each individual at one observation per second. We analyze a subset of $16$ individuals whose collars remained functional for a ten day period (419,095 time steps). In addition, in the first two days of baboon tracking, there are four group activities labeling by experts: sleeping, hanging out, coordinated progression, and coordinated non-progression~\cite{li2016adversarial}. We show later that by using only following network density as a feature to perform activity classification, we can get high accurate results of activity prediction. 

%\ivan{is there some subsetting relative to labels? In the below (commented) it mentions activity labels.}

%This dataset tracks 26 baboons over consists of 16 individuals and trajectories has a length at 419,095 time steps. The dataset recorded between August 01, 2012 and August 10, 2012 from 6 AM to 6 PM, which is nine days in total. 

%The group consist of vary types of baboon in both gender: male and female, age: junior, sub-adult, adult, and senior. This dataset covers variety of baboon's activities such as sleeping, hanging out, coordination, etc. We focus on only coordination periods to capture a leader who initiates a movement in pre-coordination time. 

\subsubsection{Fish schools trajectories}
\label{subsubsec:fish}
The movement of a fish school of golden shiners (\emph{Notemigonus crysoleucas}) are recorded by video in order to study information propagation over the visual fields of fish \cite{strandburg2013visual}. Each population contains  $70$ fish, with $10$ trained, labeled fish who are able to lead the school to feeding sites over $24$ separate coordination events. The task is to correctly identify trained fish by initiator ranking.

\subsubsection{Stock closing-price time series}
\label{subsubsec:stocks}
We collected daily closing price data for stocks listed in NASDAQ, using Yahoo! Finance.\footnote{http://finance.yahoo.com/} These time series are from January 2000 to January 2016 (4169 time-steps). We remove symbols with a large amount of missing data, leaving a total of $1443$ symbols in our dataset. Our analysis focuses on discovering large, known events and crises in an unsupervised way, and to explore initiators and sectors involved in these coordination events.

\subsection{Evaluation}
\label{EvalSec}
For synthetic datasets, we use three evaluation approaches: 
\squishlist
\item {\bf Global leadership}: For each method, we extract network and/or rank statistics over the entire time series, and report only a single aggregate initiator ranking. We compare the known ground truth ranking (used to generate the data) against the ranking of each method, reporting precision. We measure precision of identifying the true initiator, on DM, LT, IC, and INIT-1 models. For the HM model, we compare the \textit{exact} top-4 ranking against the ground truth (order matters); The evaluation is the same for CM and INIT-4 models, except the \textit{exact} top-4 ranking constraint is relaxed (i.e. we compare top-4 \textit{sets}).  

\item {\bf Local leadership}: For evaluation data in this case, we use the ground truth ranking for each local coordination event, and the time intervals of each event. We report average precision over each discovered pre-coordination interval. We evaluate the EM model using this approach. We report only the FLICA result, since it is the only method capable of producing local ranking. 

\item {\bf Initiator leadership}: For each coordination event, we measure the initiator of coordination event before coordination occurs (e.g. in the pre-coordination interval). This individual may not be highly ranked after coordination (see Fig.~\ref{fig:BaboonRankOrder}). We report the precision of global leadership considering only the pre-coordination intervals. Since only FLICA identifies pre-coordination intervals, we compare against other methods' global leadership. This evaluation demonstrates that global leadership is distinct from coordination initiation. We evaluate DM-S and HM-S models using this approach.
\squishend

\subsection{Compared leadership methods}
\begin{table}[h]
\caption{Time complexities of leadership inference methods where $n$ is a number of time series, $\omega$ is a time window, and $t^*$ is a length of time series. }
\label{MedCompTB}
\begin{center}
\begin{tabular}{ | c | c | c |} 
\hline
{\small Method }& {\small Input } &  {\small Time complexity }\\ \hline
{\small FLICA }&  {\small Time series } &  {\small $\mathcal{O}(n^2 \times t^* \times \omega )$ }\\ \hline
{\small FLOCK \cite{andersson2008reporting}} & {\small Trajectory } &  {\small $\mathcal{O}(n^2 \times t^* )$ }\\ \hline
{\small LPD \cite{kjargaard2013time}} & {\small Time series } &  { $\mathcal{O}(n^2 \times t^* \times \omega^3)$ }\\ \hline
{\small IM \cite{kempe2003maximizing}}& {\small Network }  &  {\small $\mathcal{O}(n^2 \times t^* )$ }\\ \hline
{\small Copula-Granger \cite{liu2012sparse}}& {\small Time series }  &  {\small $\mathcal{O}(n \times (\omega\times t^* )^2)$ }\\ \hline
\end{tabular}
\end{center}
\end{table}

We demonstrate the performance of our framework by comparing with previous works on influence and leadership  \cite{andersson2008reporting,kjargaard2013time,kempe2003maximizing} as well as creating the Granger-Causality framework based on the work by Liu et al.~\cite{liu2012sparse} to illustrate the potential of using Granger Causality to infer leaders in time series. These methods can infer only global initiator ranking, while our proposed framework (FLICA) can detect individual coordination events, handles switching initiator, and performs leadership model classification. Therefore, we use the global leadership identification task to compare FLICA's performance with the prior works. We report the best results under varying parameters for competing frameworks. The time complexity of each method is shown in Table~\ref{MedCompTB}.

First, the FLOCK model \cite{andersson2008reporting} identifies leaders who move toward the norm direction vector of the group and also in the front of the group. Second, LPD \cite{kjargaard2013time} creates an aggregate `following' network from time-lag features. A node is scored by breadth-first traversal on reversed `following' edges. Visited neighbors' contribution is inverse-proportional to the geodesic distance. For the purposes of our simulation, we use sliding Euclidean distance alignment (e.g. analogous to cross-correlation) because LPD does not scale to the size of our simulations under DTW (see Table \ref{MedCompTB}). Finally, for influence maximization (IM), we use the independent cascade model for the $1$-seed selection problem \cite{kempe2003maximizing}, on the network derived from \cite{andersson2008reporting}. The network describes the probability of any individual $A$ sharing the same direction as $B$, and in the front of $B$.

For the Granger Causality method, we used the Copula-Granger approach that can be found in \cite{liu2012sparse} to infer a causal network. Then, we convert the  causal network  to be a following network by designating $X$ follows $Y$ if the weight of $Y$ Granger causes $X$ is larger than  the weight of $X$ Granger causes $Y$. 

To make leadership comparison possible, we report the global leadership rank ordered list for each method as follows. First, we create rank order lists for FLICA under PageRank. The FLOCK model, however, does not have the explicit ranking score, so we rank individuals based on decreasing time duration of leadership. Third, LPD assigns individuals with higher scores a higher rank. Finally, since IM uses the probabilistic network of influence, we construct the realization of this influence network. A node influences any node to which it has a directed path in the realized network. We rank individuals based on the expectation of nodes influenced by that node over 1000 realized networks. Lastly, for the Copula-Granger leadership framework, we use PageRank to evaluate the leadership ranking on the following network we created from the causal network.

\subsection{Sensitivity analysis}
\label{sec:SenAnalysis}
Typically, real-world datasets are noisy. The high degree of noise can affect results of leadership inference. However, the effects of noise on leadership inference is unclear.  Moreover, in our leadership framework, the main parameter is the time window $\omega$. Using the wrong value of time window may affect the results as well.  Hence, in this section, we consider the approach to measure the robustness of our framework.

\subsubsection{Support of faction leading}

\begin{figure}[ht!]
\centering
\includegraphics[width=0.6\columnwidth]{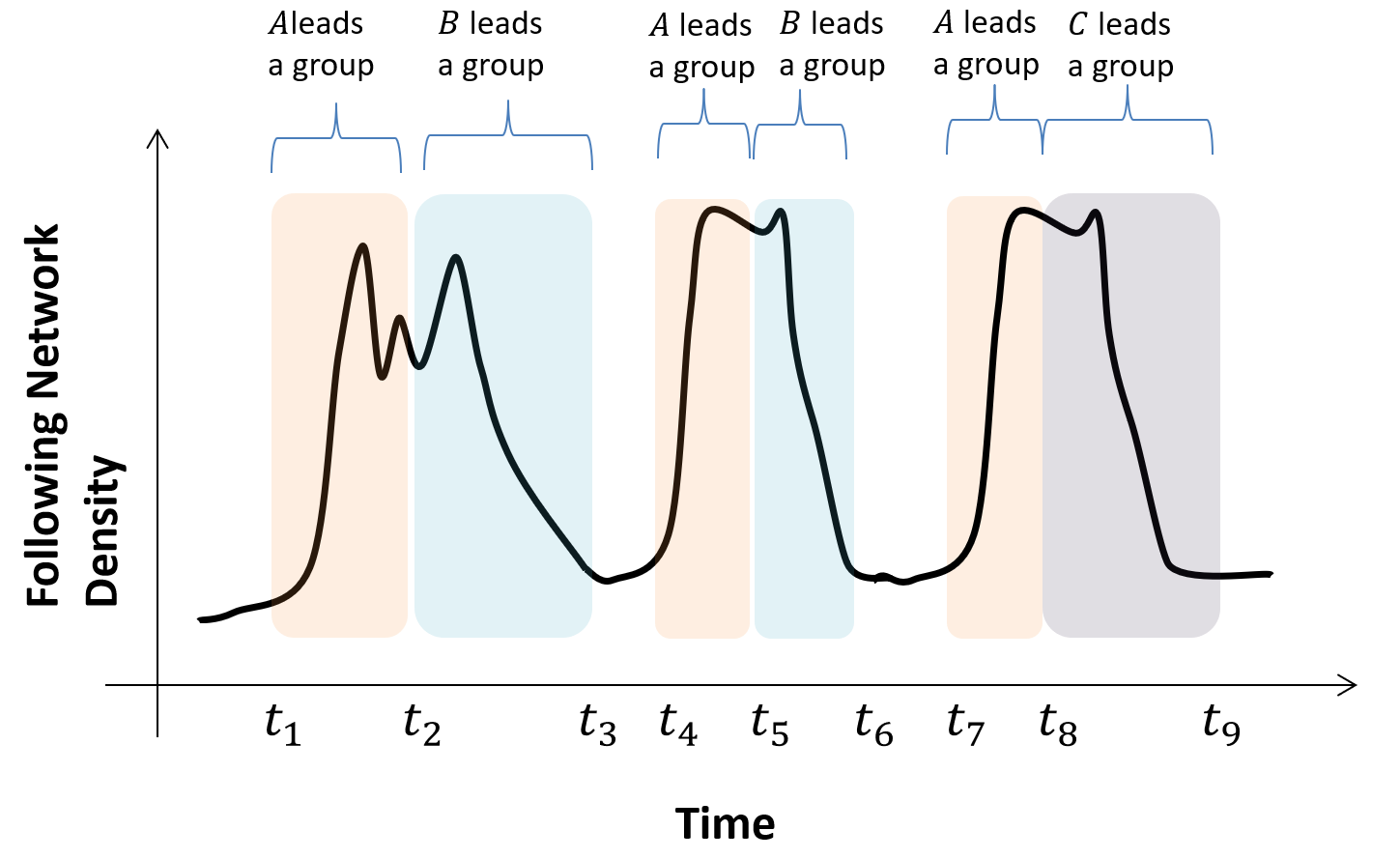}
\caption{An example of time series of following-network density. There are three coordination events where individual $A$ leads a group in first half of all coordination events while $B$ leads a group after $A$ for first and second coordination events. The third coordination event has $C$ leads the second half of the event. }
\label{fig:CoorEventSup}
\end{figure}

To measure the accuracy of initiator inference, we use a support value of being a leader of factions for each individual. A faction interval of $L$ is defined as a sub-coordinated interval within a coordination event such that $L$ leads a group. Given $\mathcal{F}=\{I_L\}$ is a set of faction intervals  where $I_L$ is a faction interval lead by $L$ (The consecutive time interval that $L$ leads the group) and a time window $\omega$, a coordination event $E_k$ is defined to be a combined interval of consecutive faction intervals. Specifically, if a faction interval $I_i$ finishes at time $t'_i$ while $I_j$ starts at time $t_j \leq t'_i+\omega$, then both $I_i$ and $I_j$ factions are in the same coordination event. 

\begin{definition}[Coordination event] 
Let $\mathcal{F}=\{I\}$ be a set of faction intervals and $\omega$ be a time window. A coordination event $E_k=[t_1,t_2]$ is a combined interval of consecutive faction intervals from $\mathcal{F}$.  Any faction interval $I_i=[t_i,t'_i]$ that occurs before another faction $I_j=[t_j,t'_j]$ are in the same coordination event if $t'_i<t_j\leq t'_i+\omega$.
\label{CoordinationEventDef}
\end{definition}

Given $\mathcal{E}=\{E_k\}$ is a set of coordination events and $\mathcal{F}_L=\{I_i\}$ is a set of faction intervals lead by $L$, the support value of an individual $L$ leading factions is defined as follows.

\begin{equation}
\mathrm{sup}(L)= \frac{\left |{\{ E |    E \in  \mathcal{E}, \exists I\in \mathcal{F}_L, I \subseteq E\}}\right | }{|\mathcal{E}|}
\label{eq:LeaderFactioSup}
\end{equation}

The support value, $\mathrm{sup}(L)$, tells us the level of consistency that $L$ happens to initiate its faction for any coordination event. If $\mathrm{sup}(L)\approx 1$, then it means $L$ always initiates factions when coordination events occur. In contrast, if  $\mathrm{sup}(L)\approx 0$, then there is a low chance that $L$ initiates any faction.

Moreover,  we can calculate a confident value of having $A$'s and $B$'s factions in the same coordination event given that $A$'s faction occurs at the event as follows. 

\begin{equation}
\mathrm{Conf}(B|A)= \frac{\left |{\{ E |    E \in  \mathcal{E}, \exists I_i \in \mathcal{F}_A,  \exists I_j \in \mathcal{F}_B, I_i,I_j \subseteq E\}}\right | }{\left |{\{ E |    E \in  \mathcal{E}, \exists I\in \mathcal{F}_A, I \subseteq E\}}\right |}
\label{eq:LeaderFactioConf}
\end{equation}

Fig.~\ref{fig:CoorEventSup} shows the toy example of coordination events in the form of density time series of following network. $\mathcal{E}=\{[t_1,t_3],[t_4,t_6],[t_7,t_9]\}$ is a set of coordination events. A set of faction intervals lead by $A$ is $\mathcal{F}_A=\{[t_1,t_2],[t_4,t_5],[t_7,t_8]\}$, a set of faction intervals lead by $B$ is $\mathcal{F}_B=\{(t_2,t_3],(t_5,t_6]\}$, and  a set of faction intervals lead by $C$ is $\mathcal{F}_C=\{(t_8,t_9]\}$. The support values of $A,B$ and $C$ are $\mathrm{sup}(A) = 1$, $\mathrm{sup}(B) = 2/3$, and $\mathrm{sup}(C) = 1/3$ respectively. The confident values of having $B$'s faction within a coordination event that has $A$'s faction is $\mathrm{Conf}(B|A)=2/3$. But $\mathrm{Conf}(A|B)=1$ and $\mathrm{Conf}(A|C)=1$. 

\subsubsection{Simulation and accuracy measure}

\squishlist
\item {\bf Initiator inference}: To measure the performance of framework vs. noise w.r.t. the initiator inference task, we use simulation datasets of movement time series within 2-dimensional space. A simulation dataset contains trajectories of individuals that have multiple coordination events (e.g. Fig.~\ref{fig:CoorEventSup}). Each type of simulation dataset contains different type and level of noises. The task is to predict a support of each initiators as well as confident values. The performance of framework is determined based on the error between ground truth and predicted values of support and confident measures. If the framework performs well, the predicted support and confident value of initiators should be close to the ground truth. 

\item {\bf Coordination interval inference}: To measure the performance of framework vs. noise w.r.t. the coordination interval inference task,  we use simulation datasets of movement time series within 2-dimensional space. we compared the predicted faction intervals of each initiators to the ground truth directly. The result of comparison are reported in the form of true positive, false positive, and false negative values. True positive will be counted at time step $t$ when a predicted and ground truth faction interval of initiator $L$ occurs at time $t$. False positive will be counted when the framework predicts that a time step $t$ is within $L$'s faction but it is not. False negative will be counted when the framework predicts that a time step $t$ is not within $L$'s faction but it is.

\squishend

\subsubsection{Position and Direction noises}
To measure the robustness of framework w.r.t. two tasks above, we consider two types of noises: position noise and direction noise. For a direction noise, instead of moving to a target direction at degree $D$ compared to $X$-axis, an individual moves toward a direction $D+a$. The direction noise $a$ is drawn randomly from a normal distribution with zero mean and $\gamma$ standard deviation. For position noise, suppose $(x,y)$ is the next position that an individual should move to, with position noise, the actual position that the individual moves is $(x+b_1,y+b_2)$.  The position noise $b_1,b_2$ are drawn randomly from a normal distribution with zero mean and  $\beta$ standard deviation. 

\subsubsection{Time window sensitivity}

A time window parameter $\omega$ is the main parameter of our leadership framework. we report the performance of framework when the time window is vary from the optimal time window. If the framework is robust, then it should perform well even when the time window value is set significantly different from the optimal value.

\section{Results}

\subsection{Identifying leaders}
In each simulation, we have the label of the true initiator(s). For each of the simulation trials, our method identifies the `initiator' and `rank ordered lists' (see Section~\ref{EvalSec}). We set a window size $\omega$ by the TWIN heuristic \cite{sulo2010meaningful} on the network density, window shift size $\delta = 0.1\omega$, and the $\lambda$ threshold at the mean of the network density time series $\mathrm{d}(t)$.  

Table \ref{SimPredictTB} reports precision on PageRank rank ordered lists over all synthetic model simulations.  We compare against previous works--FLOCK \cite{andersson2008reporting}, IM \cite{kempe2003maximizing}, LPD \cite{kjargaard2013time}, and Copula-Granger Causality Inference  models~\cite{liu2012sparse}--which produce a single ranking over the entire trial. 
%We also provide a subset of time-steps to these competing methods, using the coordination event detection of FLICA, and initiator ranking using the previous methods only on these timesteps (Table \ref{SimPredictTB}, columns 8-10).    

%to these previous  coordination event detection from FLICA   FLICA to provide The 8th-10th columns are the results from exiting methods which aggregate only pre-coordination intervals to identify a leader. 

The white rows in Table \ref{SimPredictTB} report precision of leadership identification for a fixed initiator across all coordination events  (global leadership). Gray rows report precision of initiator leadership where leaders change between pre-coordination and coordination intervals in the event (DM-S, HM-S), or precision of local leadership where the initiator changes per coordination event (EM). The rows labeled `Top-4' report precision in identifying any of the multiple unordered initiators (CM, INIT-4) or precision \emph{for the correct hierarchical order} (HM, HM-S).

On the white rows, FLICA is robust across all simulation models, while FLOCK,  IM, and LPD perform well other than on INIT-4 simulations (e.g. with multiple initiators). However, in gray rows (``initiator switching'') previous methods fail almost completely since they are unable to detect leadership prior to coordination. When the coordination state is more prevalent than the pre-coordination decision-point, ranking will favor an individual who happens to lead the dynamics in the coordination state (but may not have initiated the state). For Copula-Granger framework, it can infer correct initiators with high accuracy in Dictatorship model, while it fails for the most of models except INIT-4. This indicates that the Copula-Granger approach has a potential to infer leaders even though it is not designed to perform leadership inference.

The row reporting EM results is a special case of precision. Because we know each coordination event has a unique initiator, ranking individuals across all coordination events will fail. Instead, we report precision in identifying the initiator of \emph{each} coordination event. Since previous work generates only aggregate rankings, precision for these methods are not reported.  

%The (+FL) columns in Table \ref{SimPredictTB} show that we can leverage coordination event identification with FLICA to make previous work competitive in the leader switching case (+FL columns, grey rows). This procedure also allows us to report precision for EM simulations under these ranking models, which is also competitive under FLOCK and IM. 

\begin{table}
\caption{Precision of leadership identification on simulation models. (* indicates the $std\geq 0.1$).}
\label{SimPredictTB}
\begin{center}

\begin{tabular}
%{ |>{\centering\arraybackslash}p{2.8cm}  |>{\centering\arraybackslash}p{1.00cm}|>{\centering\arraybackslash}p{1.00cm}|>{\centering\arraybackslash}p{1.00cm}|>{\centering\arraybackslash}p{1.00cm}|}  
{| c | c | c | c | c | c | }
\hline
%\backslashbox{Models}{Methods} & FL-PCH & FLOCK & IM & LPD\\ 
Models/Methods & FLICA & FLOCK & IM & LPD & Copula-Granger\\ 
\hline

{\small DM	} & {\small\bf 1	} & {\small\bf  1 }& {\small\bf  1 }& {\small\bf 1 } & {\small\bf 0.97* }\\ \hline 
{\small HM (Top-4)	} & {\small\bf  1	}  & {\small  0.25 }& {\small\bf  1 }& {\small\bf 1 }  & {\small 0.55* }\\ \hline 
{\small $LT$	} & {\small\bf 0.99	} & {\small  0.98* }& {\small  0.99* }& {\small 0.93* } & {\small 0.07* }\\ \hline 
{\small $IC$	} & {\small\bf 1	} & {\small\bf  1 }& {\small\bf  1 }& {\small 0.99 } & {\small 0.45* } \\ \hline 
{\small CM (Top-4)	} & {\small\bf 1	} & {\small\bf  1 }& {\small\bf  1 }& {\small 0.99 } & {\small 0.69* }\\ \hline 
{\small INIT-1	} & {\small\bf 1	} & {\small\bf 1 }& {\small\bf 1 }& {\small\bf 1 } & {\small 0.24* }\\ \hline 
{\small INIT-4 (Top-4)	} & {\small\bf 0.74*	}  & {\small 0.35* }& {\small 0.51* }& {\small 0.21* } & {\small\bf 0.91* }\\ \hline

\rowcolor{LightGray}{\small DM-S} & {\small\bf 1	}& {\small 0 }& {\small 0.02* }& {\small 0.25* } & {\small 0.37* } \\ \hline 

\rowcolor{LightGray}
{\small HM-S (Top-4)	} & {\small\bf 1	} & {\small 0 }& {\small 0.5 }& {\small 0.51 } & {\small 0.31* }\\ \hline 

\rowcolor{LightGray}
{\small EM	} & {\small\bf 0.92	} & {\small - }& {\small -}& {\small - } & {\small - }\\ \hline 
\rowcolor{LightBlue}
{\small Random	} & {\small 0.01	} & {\small 0 }& {\small 0.01 }& {\small 0.17* } & {\small 0.02* }\\ \hline 
\end{tabular}
\end{center}
\end{table}

\subsection{Case study: trained initiators in fish schools}

We identify the top-$k$ global initiators of the fish school trajectory dataset (see Section \ref{subsubsec:fish}), where we have the labels of `trained' individuals expected to lead the school to feeding sites. Table \ref{FishPredictTB} reports precision of identifying trained fish as initiators over $24$ trials. The Initiator column is precision of predicting a trained fish as a global initiator. The Top-4 rank column is precision of identifying trained fish as the top-4 ranking individuals. Similar to the simulation models, FLICA performs best overall, again suggesting that dynamic following network representation captures `following'  better than other features. %Significant noise in the fish dataset causes previous methods to perform poorly. 

\begin{table}[ht]
\caption{Initiator identification precision in fish (* indicates the $std\geq 0.1$). }
\label{FishPredictTB}
\begin{center}
\begin{tabular}{ | c | c | c |} 
\hline
{\small Ranking }& {\small Initiator} & {\small Top-4 rank} \\ \hline
{\small FLICA }& {\small \bf 	0.83*} & {\small \bf 	0.61*} \\ \hline
%FL-VCH & { 	0.71} & { 	0.65} \\ \hline
%FL-PCH & { 	0.67} & { 	0.51} \\ \hline
{\small FLOCK \cite{andersson2008reporting}  }& {\small 	0.0} & {\small 	0.0} \\ \hline
{\small IM \cite{kempe2003maximizing} }& {\small 	0.0} & {\small 	0.02} \\ \hline
{\small LPD \cite{kjargaard2013time} }& {\small 	0.17*} & {\small 	0.18*} \\ \hline
{\small Copula-Granger~\cite{liu2012sparse} }& {\small 	0.13*} & {\small 	0.10*} \\ \hline
\end{tabular}
\end{center}
\end{table}

\subsection{Case study: finding ``initiators" of stock market events} 
We apply our leadership framework to stock market closing price data of the NASDAQ index. An `initiator' in this context measures the extent that a stock increases or decreases in value before a large group of other stocks (e.g. a coordinated group). We apply the framework without any special consideration to the domain, only to qualitatively validate that we can discover known, large events. 

\begin{figure}[ht]
\centering     %%% not \center
%\captionsetup[subfigure]{labelformat=empty}
\subfigure{\label{fig:stocks-exA}\includegraphics[width=.9\columnwidth]{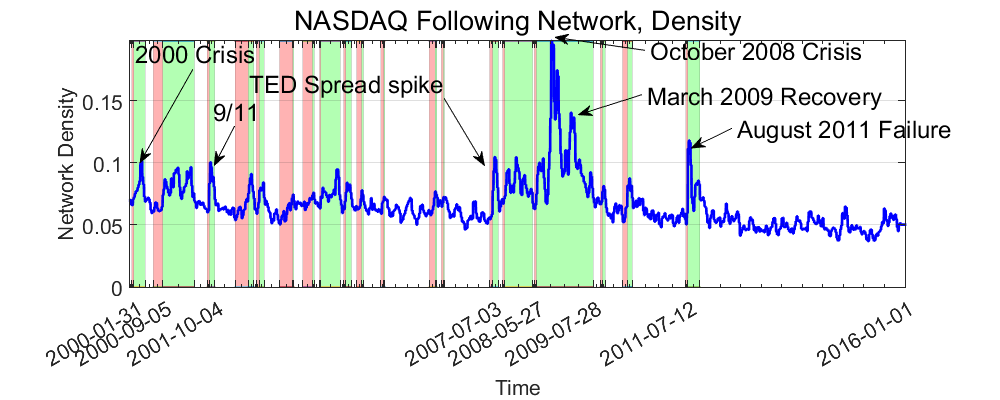}}
\subfigure{\label{fig:stocks-exB}\includegraphics[width=.9\columnwidth]{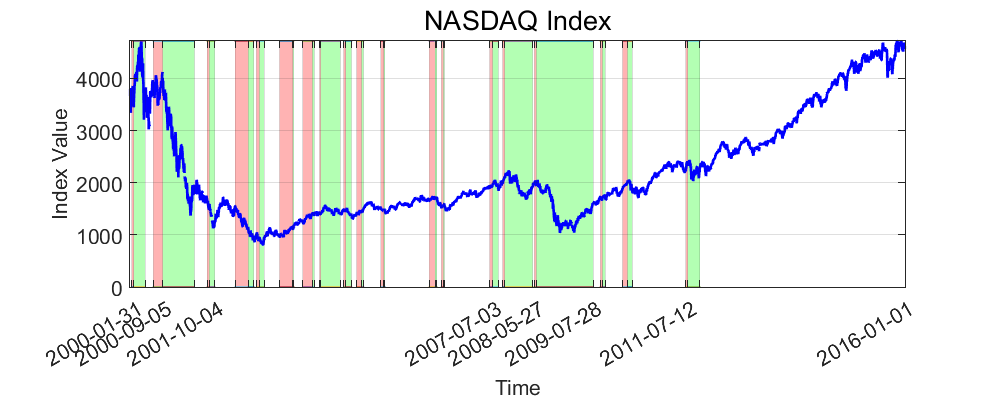}}
\caption{(Top) NASDAQ `following' network density and (Bottom) NASDAQ index value. Pre-coordination and coordination intervals are shown in red and green, respectively. The framework detects many known events in financial data (labeled above). Many of these events are not reflected in the NASDAQ index. }
\label{fig:stockmarket}
\end{figure}

Fig.~\ref{fig:stockmarket} shows the network density of the inferred `following' network over time, where we discover coordination events with $\lambda$ threshold at the 75th percentile of the network density time series. Pre-coordination and coordination intervals are shown in red and green, respectively. We find significant economic events such as the 2000 tech collapse, and 9/11. More interestingly, we discover known events which are reflected in the network density signal but not the NASDAQ index. For example, we discover a technical econometric event, where the ``TED Spread'' (a surrogate of national credit risk) begins fluctuating in July 2007, and a small market failure in August 2011. Matching our intuition, the top-ranked companies in the coordination event associated with the year 2000 collapse are primarily in IT and semiconductors, including eBay and SanDisk in the top 10.

\begin{figure}[ht]
\centering
\includegraphics[width=1\columnwidth]{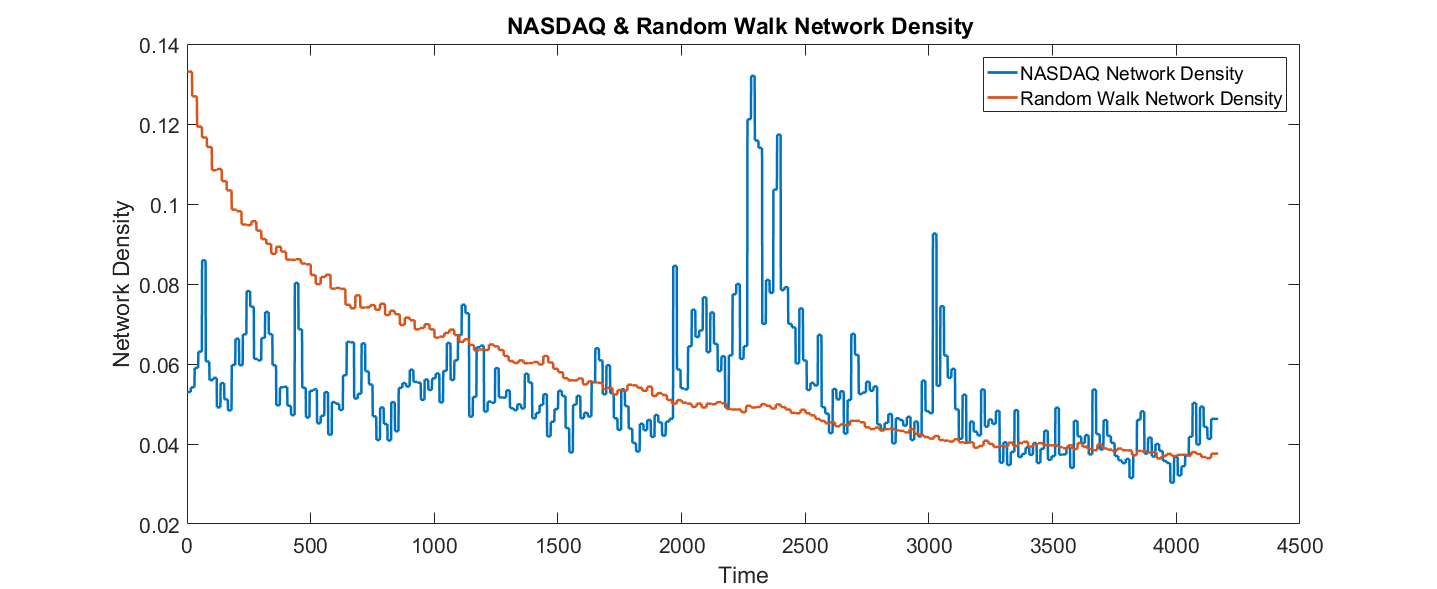}
\caption{Comparison between time series of network density  that generated from NASDAQ time series and from random walk time series.  }
\label{fig:RandWalk}
\end{figure}

For the sanity check, we provided the results of the comparison between the following network density of NASDAQ stock market and a random walk one in Fig.~\ref{fig:RandWalk}. We generated random-walk time series from the original NASDAQ closing price time series. Both original and random-walk versions shared the same distribution of difference between time steps, length, and the number of time series. 

For each time series of closing price $X$, we inferred the distribution $\mathcal{D}_X$ of the differences between the time steps. We created random-walk time series $\hat{X}$ that starts as the same price as the original time series, then we updated the next value of $\hat{X}$ by normally sampling a different value from $\mathcal{D}_X$. Hence, both $X$ and $\hat{X}$ share the same $\mathcal{D}_X$.

We found that our following network density of NASDAQ is different from the random-walk one (Fig.~\ref{fig:RandWalk}). Moreover, random-walk network density does not have any coordination events. This indicates that our network density can tell the difference between random-walk time series and the actual dataset that contains coordination events.

\subsection{Case study: baboon activity classification by following network density} 
\label{sec:ActCls}

\begin{figure}[ht]
\centering
\includegraphics[width=.8\columnwidth]{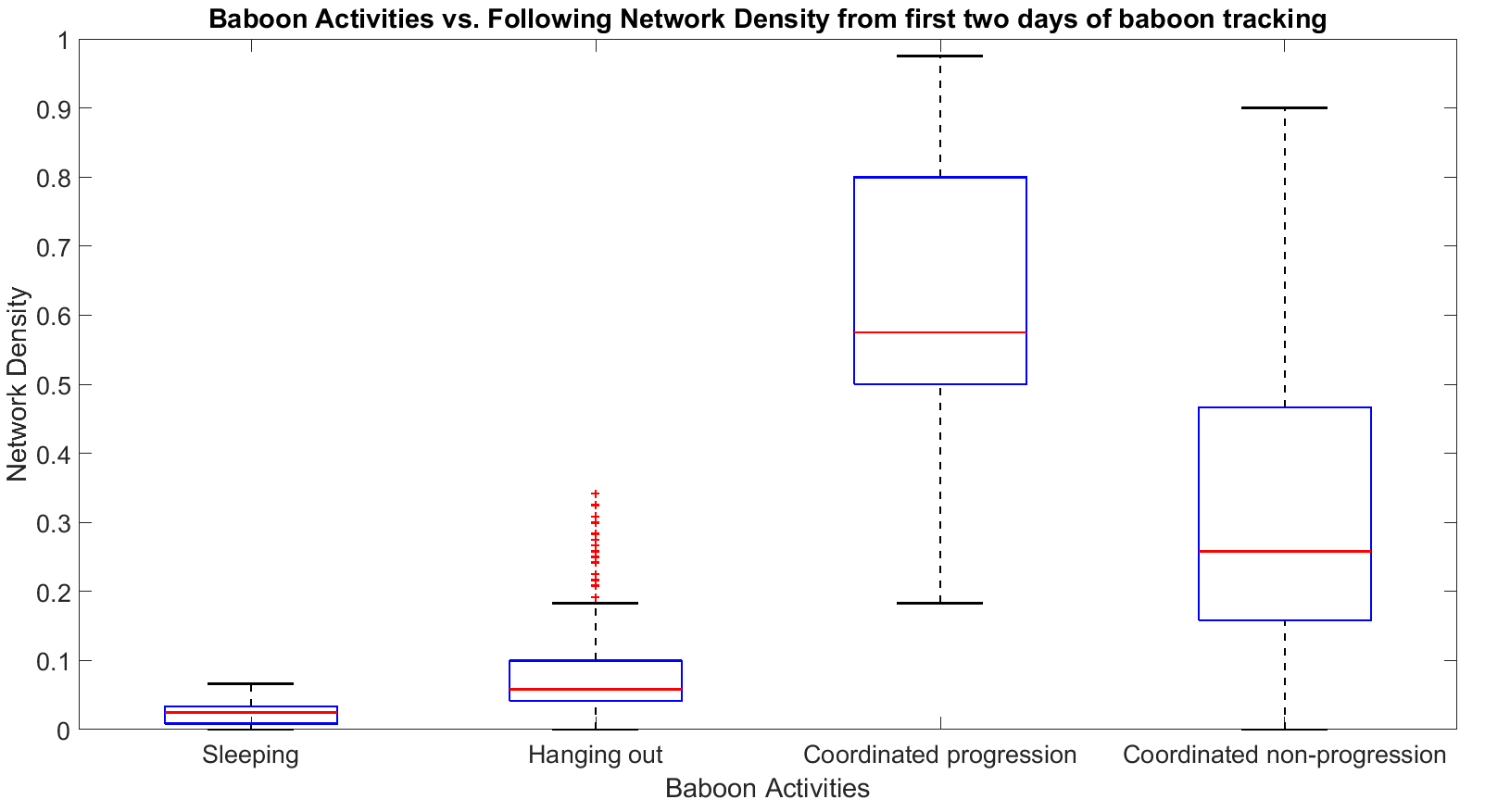}
\caption{Baboon activities from first two days of tracking vs. following network density}
\label{fig:BaboonActRes}
\end{figure}

In this section, we demonstrate that our coordination events, which defined by following network density time series, corresponding to coordinated progression activity labeled by experts in baboon dataset (see Section~\ref{subsubsec:baboon}).

For each time step, a group of baboon has an activity label as either sleeping, hanging out, coordinated progression, or coordinated non-progression. The group of baboons is considered to have a sleeping label when they are at their home tree to sleep. Hanging out activity happens when baboons stay around the same place without moving far away from the group. Coordinated progression is when a group of baboons having a strong coordinated movement to somewhere. Lastly,  in coordinated non-progression, a group has a weaker coordinated movement than coordinated progression.  

Fig.~\ref{fig:BaboonActRes} shows the distribution of network density for each activity. Sleeping and hanging out activities have low values of network density, which implies that the group rarely has following relations.   On the contrary, coordinated progression and coordinated non-progression have high values of network density on average, which is similar to our coordination intervals. This result illustrates that a following network density is informative with respect to the task of activity classification. 

In the dataset, there are two days of baboon activity labels. We used the first day data to train a classifier to predict the second day activities as well as using the second day to train a classifier to predict the first day activities. We used  Linear discriminant analysis (LDA) as our classifier and used only network density as a feature. We compared our result with Adversarial Sequence Tagging (AST)~\cite{li2016adversarial}, with is the state-of-the-art method that performed activity labeling classification in the same dataset.  Our aim is to show that a following network density is informative enough to make a simple classifier performs better than the state-of-the-art classifier method that used 24 features in the group activity classification task.

\begin{table}[ht]
\caption{Activities classification prediction accuracy in first two days of baboon data. }
\label{BaboonAccPredictTB}
\begin{center}
\begin{tabular}{ | c | c | c |} 
\hline
{\small Method }& {\small Baboon Day 1} & {\small Baboon Day 2} \\ \hline
{\small Linear discriminant analysis (LDA) + following network density }& {\small \bf 	87.20\%} & {\small \bf 	70.82\%} \\ \hline
{\small Adversarial Sequence Tagging (AST)~\cite{li2016adversarial} + 24 features  }& {\small 77.30\%} & {\small 	69.22\%} \\ \hline

\end{tabular}
\end{center}
\end{table}

The classification result is shown in Table~\ref{BaboonAccPredictTB}. The accuracy results were calculated by the complement of hamming loss the same as ~\cite{li2016adversarial}. By using only following network density as a feature, our simple classifier performs better than AST in both days.

\subsection{Leadership model classification}

%============== begin figure ===============
\begin{figure}[ht]
\centering
\includegraphics[width=1\columnwidth]{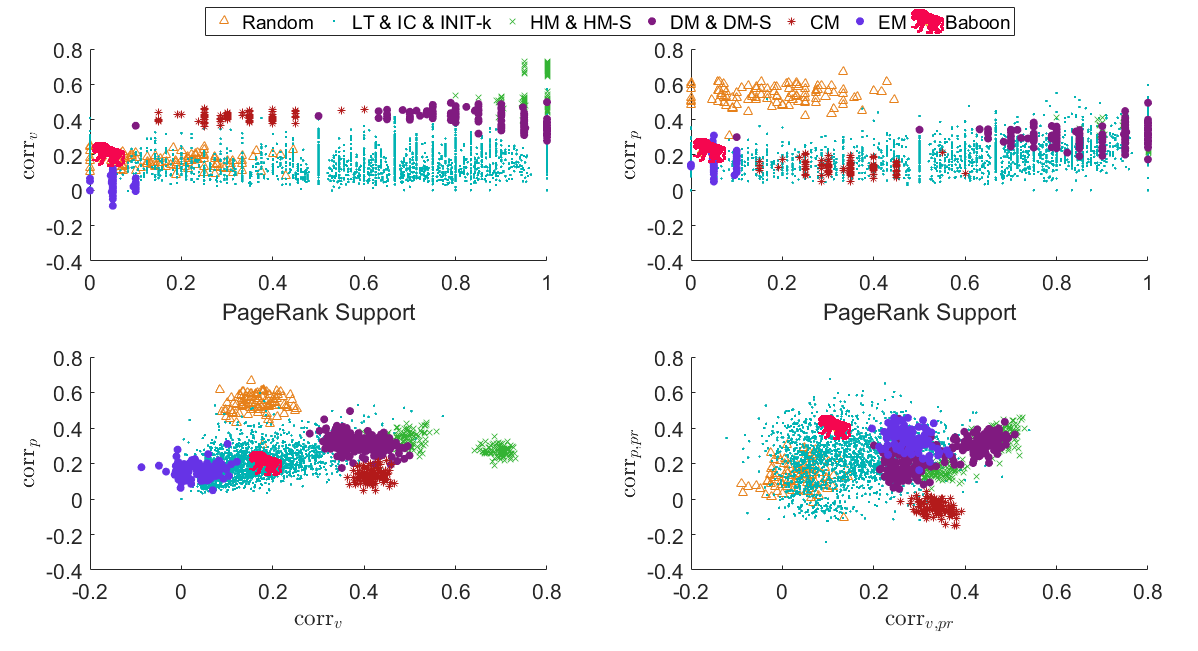}
\caption{Comparison of feature spaces of leadership model classifications on simulations and real data}
\label{fig:ModelFeaturesMainNew}
\end{figure}
%============== end figure ===============

Recall, that we proposed several initiator rankings and ranking correlations (Section \ref{subsec:features-rank}). Here, we do leadership model classification on each simulation trial using the proposed features derived from the rank correlations: $\mathrm{corr}_{p}$, $\mathrm{corr}_{v}$, $\mathrm{corr}_{p, pr}$, $\mathrm{corr}_{v, pr}$ and $\mathrm{sup}_{pr}$. A classifier takes these features and produces a leadership model label, one per trial of the simulation model in the evaluation hold-out. We use 10-fold cross validation on Random Forests \cite{ho1998random} over the 2700 total trials and report mean precision and recall across folds. Table \ref{FeaturesLCls} reports the classification results for each simulation model. We combine some models into a shared label because they share similar characteristics when we project them into our feature spaces. For example, DM, and DM-S models always have high $\mathrm{corr}_{v}$ but low $\mathrm{corr}_{p}$. 
 
Fig.~\ref{fig:ModelFeaturesMainNew} visualizes sub-spaces of the full feature-space. Fig.~\ref{fig:ModelFeaturesMainNew} (Top) shows the maximum support ($\mathrm{sup}_{pr}$) over all individuals for this trial vs. the $\mathrm{corr}_{v}$ (the rank correlation between global and local VCH ranking) and $\mathrm{corr}_{p}$. The $\mathrm{sup}_{pr}$ axis (x-axis) describes how `dictatorial' (e.g. consistent) the leadership is across coordination events. DM therefore has high support, while EM (distinct leaders per coordination event) has low $\mathrm{sup}_{pr}$. The $\mathrm{corr}_{v}$ and $\mathrm{corr}_{p}$ axes describe consistency between local and global convex hull rankings. HM has high velocity ranking because leaders accelerate in a consistent sequence, yielding consistent individuals movement outside of the VCH in the previous time step. The random model produces high $\mathrm{corr}_{p}$ because relative positions within the group are somewhat consistent. Therefore, a consistent set of individuals expand the PCH from the previous time step.

Fig.~\ref{fig:ModelFeaturesMainNew} (Bottom-Right) reports the mean rank correlation between PageRank rank ordering, against PCH and VCH ranking \emph{in each coordination event}. At the origin (0, 0), ranking from the inferred `following' network is uncorrelated with time series feature rankings in position or velocity. Following our intuition, the Random simulation has the lowest cross-domain feature correlation, while DM and HM have highest correlation between these domains. As the simplest simulations, DM and HM both dictate that leaders will have regular position (e.g. the front of the group), or velocity (accelerating in sequence before others). Simulations such as CM, LT, IC have indirect relationships between relative position and velocity vs. the following network ranking.

\subsubsection{Baboon leadership model characterization} 
A key aspect of our simulation modeling is that we can characterize real datasets according to how they map into these  feature-spaces, compared to synthetic models. We compute each rank correlations over high-confidence baboon events, labeled  ``Baboon'' in Fig.~\ref{fig:ModelFeaturesMainNew}, thresholded at the $99$th percentile of density. We observe that within different sub-spaces, the baboon ranking is similar to Random or Linear Threshold, and has low maximum support for global vs. local rank correlation features (e.g. $\mathrm{corr}_{p}$). We see this rank correlation between both cross-domain axes (Fig.~\ref{fig:ModelFeaturesMainNew} (Bottom-Right)). This suggests that in aggregate, baboon leadership is heterogeneous and context-driven, though overall closer to the Linear Threshold influence model (as biologically expected).  This analysis provides a strategy for hypothesis testing and generation on contrasting time-scales and sub-spaces.

	%====== Model selection  ==================
\begin{table}[h!]
\caption{Random forest classification of synthetic leadership models using proposed features}
\label{FeaturesLCls}
\begin{center}
\begin{tabular}{ | c | c | c | c | c |} 
\hline
{\small Model }& {\small Precision} & {\small Recall} & {\small F1-score} \\ \hline
{\small DM, DM-S} & {\small 0.86	} & {\small 0.80	} & {\small 0.81} \\ \hline
{\small HM, HM-S } & {\small 0.69	} & {\small 0.98	} & {\small 0.80} \\ \hline
{\small LT, IC, INIT-k} & {\small 0.99	} & {\small 0.97	} & {\small 0.98} \\ \hline
{\small CM	} & {\small 0.75	} & {\small 0.94	} & {\small 0.80} \\ \hline
{\small EM	} & {\small 1	} & {\small 0.54	} & {\small 0.64} \\ \hline
{\small Random	} & {\small 0.98	} & {\small 0.95	} & {\small 0.97} \\ \hline
\end{tabular}
\end{center}
\end{table}

\subsection{Following network density perturbation in baboon and fish data}

In this section, we provided the results of the network density changes when individuals are removed from the datasets, either the high-rank ones or (uniformly) at random.

In the baboon dataset, we used leadership ranking during pre-coordination intervals to choose the top-$k$ rank individuals. In the fish datasets, we used leadership ranking to choose top-$k$ ranked individuals, which are also informed fish. For the randomly chosen individuals, we uniformly and randomly chose $k$ individuals from the population. We repeated the random choice process 100 times and report the average of these results.

\begin{table}[]
\centering
\caption{The average difference of network density between before and after removing $k$ individuals on the baboon dataset. We compare the case of removing high-rank individuals vs. random individuals. An element in the table represents the difference between the original network density and the network density after removing $k$ individuals.}
\label{table:baboonPerturb}
\begin{tabular}{c|c|c|c|c|}
\cline{2-5}
                                & \multicolumn{4}{c|}{Number of Individuals being removed from 16 individuals (\#$k$)}                                         \\ \hline
\multicolumn{1}{|c|}{}          & \#1                      & \#2                     & \#3                    & \#4                     \\ \hline
\multicolumn{1}{|c|}{High-Rank} & $-5.45\times 10^{-03}$   & $-7.37\times 10^{-03}$  & $-9.65\times 10^{-03}$ & $-10.14\times 10^{-03}$ \\ \hline
\multicolumn{1}{|c|}{Random}    & $0.057\times 10^{-03}$ & $-0.004\times 10^{-03}$ & $-0.504\times 10^{-03}$ & $0.002\times 10^{-03}$ \\ \hline
\end{tabular}
\end{table}

Table~\ref{table:baboonPerturb} shows the average difference of the network density before and after removing $k$ individuals from the baboon dataset. In the 'High-Rank' row, the network density decreases with the removal of high-ranked individuals. In contrast, in the 'Random' row, the network density is largely unaffected by the uniformly random removal of the individuals. 

\begin{table}[]
\centering
\caption{The average difference of the network density between before and after removing $k$ individuals from the fish data. We compare the case of removing high-rank individuals vs. random individuals. An element in the table represents the difference between the original network density and the network density after removing $k$ individuals.}
\label{table:fishPerturb}
\begin{tabular}{c|c|c|c|c|}
\cline{2-5}
                                & \multicolumn{4}{c|}{Number of Individuals being removed from 70 individuals (\#$k$)}                                      \\ \hline
\multicolumn{1}{|c|}{}          & \#2                     & \#4                    & \#6                    & \#8                    \\ \hline
\multicolumn{1}{|c|}{High-Rank} & $-20.26\times 10^{-04}$ & $-3.96\times 10^{-04}$ & $-2.45\times 10^{-04}$ & $-8.44\times 10^{-04}$ \\ \hline
\multicolumn{1}{|c|}{Random}    & $-0.55\times 10^{-04}$   & $-0.04\times 10^{-04}$ & $-3.90\times 10^{-04}$ & $-3.88\times 10^{-04}$ \\ \hline
\end{tabular}
\end{table}

In the fish data, Table~\ref{table:fishPerturb} shows the results of the network density difference after removing $k$ individuals.  In the 'High-Rank' row, the network density decreases with the removal of high-ranked individuals, albeit less so than in the baboon dataset. In the random case, the network density is still largely unaffected by the uniformly random removal of the individuals. 

The results above show that by removing high-rank individuals in both baboon and fish datasets, the network density decreases significantly compared to randomly removed individuals. Moreover, we found that removing high-rank individuals in the baboon dataset resulted in a larger decrease of the network density than in the fish datasets. This suggests that baboons have a stronger following hierarchy than schools of fish.

\subsection{Sensitivity analysis}
\begin{figure}[ht]
\centering
\includegraphics[width=1\columnwidth]{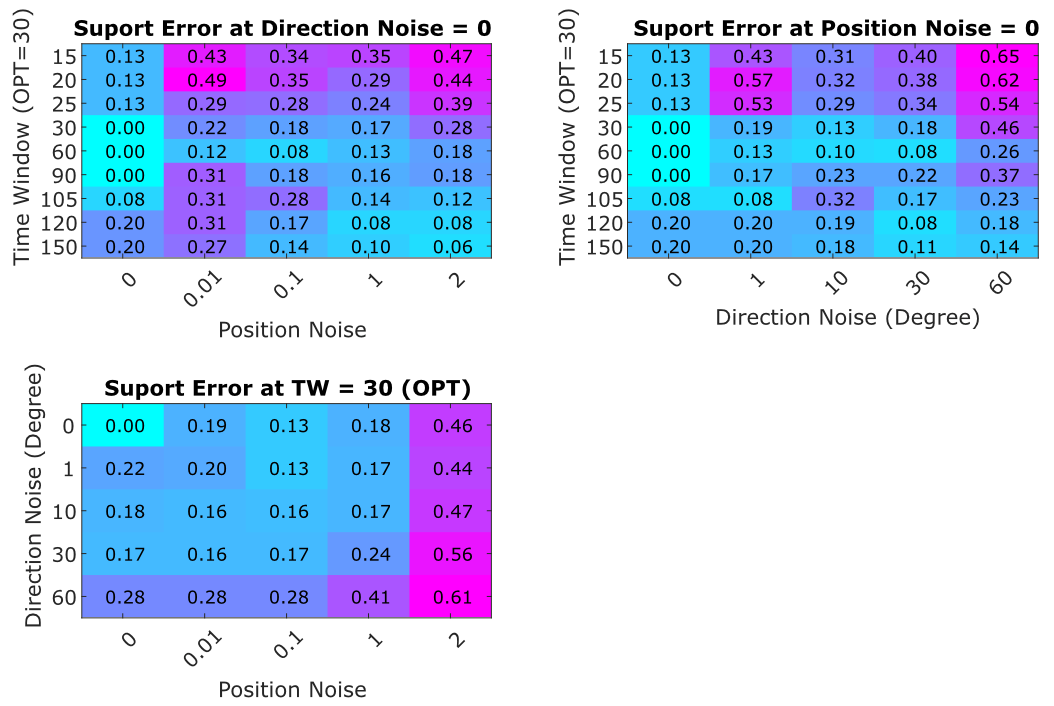}
\caption{Loss values of prediction of the initiator support  for different levels of noise and time window sizes. A lower value implies a better prediction result.}
\label{fig:SupSenAz}
\end{figure}

We conducted the sensitivity analysis based on Section~\ref{sec:SenAnalysis} to demonstrate the robustness of our framework. We used simulation datasets that have the time delay for following relations less than 30 time steps. Hence, the optimal time window $\omega$ is 30 time steps. In the initiator inference, the results of loss values of the initiator-support prediction are shown in Fig.~\ref{fig:SupSenAz}.  Each cell in these three sub figures represents a loss value which is the difference between the ground-truth and the predicted value of the initiator support (Eq.~\ref{eq:LeaderFactioSup}). In general, unsurprisingly, the loss value increases with noise.  

Fig.~\ref{fig:SupSenAz} (below) illustrates that both the position and direction noise affect the prediction performance.   In top-left and top-right plots, they show that when we set the time window below the optimal value ($\omega<30$), the loss of support inference is significantly higher than setting the time window above the optimal time window.  This suggest us to try to guess the possibly maximum time delay in datasets and avoid setting the time window parameter below this value if the ground truth regarding time delay is not available. 

\begin{figure}[ht]
\centering
\includegraphics[width=1\columnwidth]{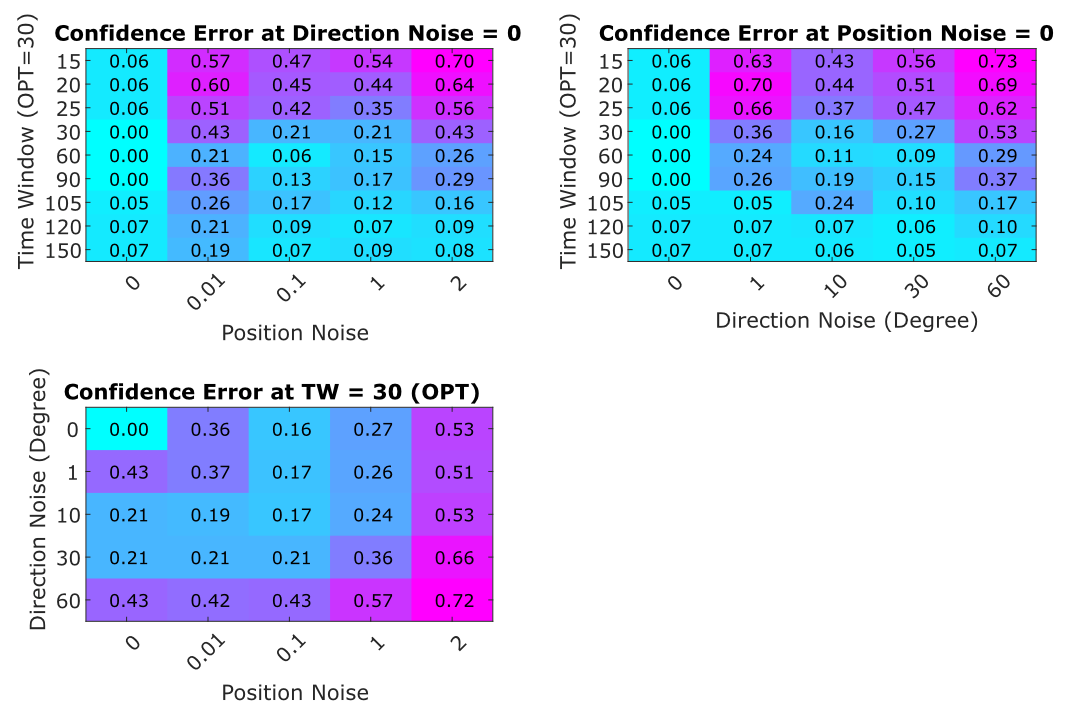}
\caption{Loss values of prediction of the initiator confidence  for different levels of noise and time window sizes. A lower value implies a better prediction result.}
\label{fig:ConfSenAz}
\end{figure}

Fig.~\ref{fig:ConfSenAz} shows the loss values of the initiator-confidence prediction. Each value in these three sub figures represents a loss value, which is a difference between the ground-truth and the predicted confidence values. Similar to the support result, higher levels  of noise results in higher loss values. Similarly, setting the time window below the optimal value also severely affects the framework performance in confidence prediction.

\begin{figure}[ht]
\centering
\includegraphics[width=1\columnwidth]{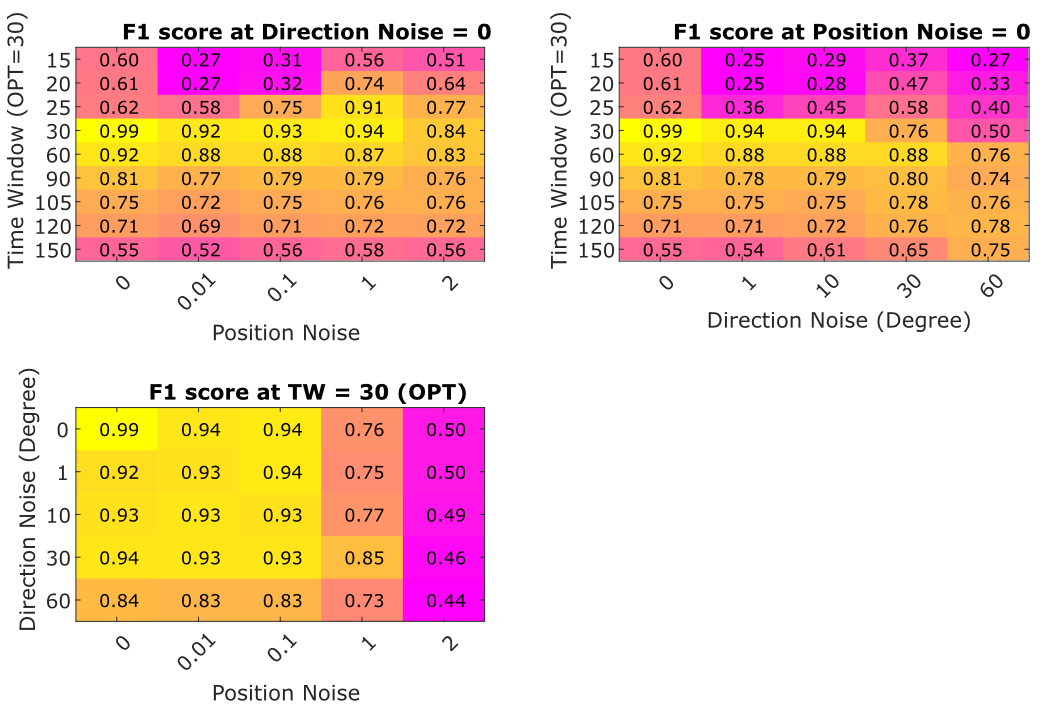}
\caption{F1 scores of coordination inference for different levels of noise and time window sizes.  A higher value implies a better prediction result. }
\label{fig:F1SenAz}
\end{figure}

In the coordination inference task, Fig.~\ref{fig:F1SenAz} shows the F1 scores of the coordination inference for different levels of noise and time window sizes. Each value in these three sub figures represents the F1 score of the prediction.  Similar to the previous results in the initiator inference, higher levels of noise reduce prediction performance. Setting the time window below the optimal value also decreases the prediction performance severely. 

In conclusion, when we increase the amount of noise in the datasets, the framework performance decreases.  Moreover, we found that if we set the time window parameter below the optimal, it severely affects the framework performance in both initiator and coordination inference tasks. In contrast, when we set the time window above the optimal value, the framework performance drops only slightly. The optimal time window is, then, a value above the possible highest time delay that two individuals can follow each other in the dataset. This suggests that the users of our framework should try to come with an educated guess of the possible values of time delays that are relevant for the application context and set the time window parameter accordingly. 
\section{Conclusions}
We narrow the gap between the biosociological view of leadership in group decision-making and the computational approaches to leadership inference.
The work presented in this paper formalizes a {\bf new computational problem}, namely \clip, and proposes the concrete, simple yet powerful, unsupervised general framework as a solution.  The framework is capable of (1) identifying events of coordinated group behavior, (2) identifying leaders as initiators of these events, and (3) classifying the type of leadership process at play. We validate the accuracy of our framework in performing all three of these tasks using 2,700 simulated datasets. Since there are no methods for local leadership inference and leadership model classification, we compared our framework with the state-of-the-art methods for global leadership identification. Our method performance is consistently competitive and its abilities go beyond other approaches in all datasets. 
%The model features we presented here can be used to distinguish different models of leadership.  
We further show that the framework can provide insights on real-world data, including data on collective animal movement and the economy. The methodology presented here is general and applicable to a wide variety of domains where coordination across many individuals or entities is observed. Moreover, our framework is highly flexible, and can easily be extended to incorporate other models of leadership or other features used in model classification, depending on the details of the system being analyzed. For reproducibility, we provide our code and simulation datasets at \cite{FLICAwebsite}.

\bibliographystyle{ACM-Reference-Format-Journals}
%%% -*-BibTeX-*-
%%% Do NOT edit. File created by BibTeX with style
%%% ACM-Reference-Format-Journals [18-Jan-2012].

\end{document}